\documentstyle{mn}
\input epsf
\newif\ifAMStwofonts
\AMStwofontstrue

\def\aj{{AJ}}                   % Astronomical Journal
\def\araa{{ARA\&A}}             % Annual Review of Astron and Astrophys
\def\apj{{ApJ}}                 % Astrophysical Journal
\def\apjl{{ApJ}}                % Astrophysical Journal, Letters
\def\apjs{{ApJS}}               % Astrophysical Journal, Supplement
\def\aap{{A\&A}}                % Astronomy and Astrophysics
\def\aapr{{A\&A~Rev.}}          % Astronomy and Astrophysics Reviews
\def\mnras{{MNRAS}}             % Monthly Notices of the RAS
\def\pasj{{PASJ}}               % Publications of the ASJ
\let\apjlett=\apjl

\ifoldfss
  \ifCUPmtlplainloaded \else
    \NewTextAlphabet{textbfit} {cmbxti10} {}
    \NewTextAlphabet{textbfss} {cmssbx10} {}
    \NewMathAlphabet{mathbfit} {cmbxti10} {} % for math mode
    \NewMathAlphabet{mathbfss} {cmssbx10} {} %  "   "    "
  \fi
  \ifAMStwofonts
    \ifCUPmtlplainloaded \else
      \NewSymbolFont{upmath} {eurm10}
      \NewSymbolFont{AMSa} {msam10}
      \NewMathSymbol{\upi}     {0}{upmath}{19}
      \NewMathSymbol{\umu}     {0}{upmath}{16}
      \NewMathSymbol{\upartial}{0}{upmath}{40}
      \NewMathSymbol{\leqslant}{3}{AMSa}{36}
      \NewMathSymbol{\geqslant}{3}{AMSa}{3E}

    \fi
  \fi
\fi % End of OFSS

\ifnfssone
  \newmathalphabet{\mathit}
  \addtoversion{normal}{\mathit}{cmr}{m}{it}
  \addtoversion{bold}{\mathit}{cmr}{bx}{it}
  \newmathalphabet{\mathbfit} % math mode version of \textbfit{..}
  \addtoversion{normal}{\mathbfit}{cmr}{bx}{it}
  \addtoversion{bold}{\mathbfit}{cmr}{bx}{it}
  \newmathalphabet{\mathbfss} % math mode version of \textbfss{..}
  \addtoversion{normal}{\mathbfss}{cmss}{bx}{n}
  \addtoversion{bold}{\mathbfss}{cmss}{bx}{n}
  \ifAMStwofonts
    \ifCUPmtlplainloaded \else
      \UseAMStwoboldmath
      \makeatletter
      \new@mathgroup\upmath@group
      \define@mathgroup\mv@normal\upmath@group{eur}{m}{n}
      \define@mathgroup\mv@bold\upmath@group{eur}{b}{n}
      \edef\UPM{\hexnumber\upmath@group}
      \new@mathgroup\amsa@group
      \define@mathgroup\mv@normal\amsa@group{msa}{m}{n}
      \define@mathgroup\mv@bold\amsa@group{msa}{m}{n}
      \edef\AMSa{\hexnumber\amsa@group}
      \makeatother
      \mathchardef\upi="0\UPM19
      \mathchardef\umu="0\UPM16
      \mathchardef\upartial="0\UPM40
      \mathchardef\leqslant="3\AMSa36
      \mathchardef\geqslant="3\AMSa3E
    \fi
  \fi
\fi % End of NFSS release 1

\ifnfsstwo
  \DeclareMathAlphabet{\mathbfit}{OT1}{cmr}{bx}{it}
  \SetMathAlphabet\mathbfit{bold}{OT1}{cmr}{bx}{it}
  \DeclareMathAlphabet{\mathbfss}{OT1}{cmss}{bx}{n}
  \SetMathAlphabet\mathbfss{bold}{OT1}{cmss}{bx}{n}
  \ifAMStwofonts
    \ifCUPmtlplainloaded \else
      \DeclareSymbolFont{UPM}{U}{eur}{m}{n}
      \SetSymbolFont{UPM}{bold}{U}{eur}{b}{n}
      \DeclareSymbolFont{AMSa}{U}{msa}{m}{n}
      \DeclareMathSymbol{\upi}{0}{UPM}{"19}
      \DeclareMathSymbol{\umu}{0}{UPM}{"16}
      \DeclareMathSymbol{\upartial}{0}{UPM}{"40}
      \DeclareMathSymbol{\leqslant}{3}{AMSa}{"36}
      \DeclareMathSymbol{\geqslant}{3}{AMSa}{"3E}
    \fi
  \fi
\fi % End of NFSS release 2

\ifCUPmtlplainloaded \else
  \ifAMStwofonts \else % If no AMS fonts
    \def\upi{\pi}
    \def\umu{\mu}
    \def\upartial{\partial}
  \fi
\fi

\begin{document}
\title[Measuring mass loss rates from satellites]
{Measuring mass loss rates from Galactic satellites}

\author[Johnston, Sigurdsson \& Hernquist]
{Kathryn V. Johnston,$^1$ Steinn Sigurdsson,$^2$ and Lars 
Hernquist$^3$ \\
$^1$ Institute for Advanced Study, Princeton, NJ 08450 \\
$^2$ Institute of Astronomy, Madingley Road, Cambridge, CB3 0HA, England \\
$^3$ Department of Astronomy and Astrophysics, University
of California, Santa Cruz, CA 95064.}

\maketitle
\begin{abstract}

We present the results of a study that uses numerical simulations to
interpret observations of tidally disturbed satellites around the
Milky Way.  When analysing the simulations from the viewpoint of an
observer, we find a break in the slope of the star count and velocity
dispersion profiles in our models at the location where unbound stars
dominate.  We conclude that `extra-tidal' stars and enhanced
velocity dispersions observed in the outskirts of Galactic satellites
are due to contamination by stellar debris from the tidal interaction with
the Milky Way.  However, a significant bound population can exist
beyond the break radius and we argue that it should not be identified
with the tidal radius of the satellite.

We also develop and test a method for determining the mass loss rate
from a Galactic satellite using its extra-tidal population.  We apply
this method to observations of globular clusters and dwarf spheroidal
satellites of the Milky Way, and conclude that a significant fraction
of both satellite systems are likely be destroyed within the next
Hubble time.

Finally, we demonstrate that this mass loss estimate allows us to
place some limits on the initial mass function (IMF) of stars in a
cluster from the radial dependence of its present day mass function
(PDMF).

\end{abstract}


\section{INTRODUCTION}

Accurate estimates of mass loss rates from the satellite galaxies and
globular clusters in orbit around the Milky Way offer the possibility
of better understanding the dynamical history of the Galaxy.  When
integrated over the entire population, such measurements would tell us
about the current accretion rate onto the Galaxy.  This in turn would
indicate whether the Galaxy has grown significantly through the
gradual process of tidal stripping and disruption of its satellites.
The direct inference of mass loss rates from individual objects would
place new constraints on detailed analytic or numerical
dynamical models of these systems.  In addition, the measurements
could be used to identify satellites which are likely to have
well-populated streams of tidal debris associated with them.

In particular, the formation, evolution and ultimate fate of the
Galactic globular clusters is a long standing puzzle in astrophysics.
Dynamically, globular clusters are clean and relatively simple
systems, yet on close inspection they reveal a wealth of complex
behaviour, including relaxation and evaporation from internal dynamical
interactions and mass loss in response to the tidal field of the Milky
Way \cite{me97,el87}.  Ideally
we would like to be able to assess the importance of each of these
processes for the evolution of individual globular clusters from
observations.

Of particular interest for mass-segregated, relaxed clusters is the
differential mass loss of stars due to the different relative
densities of low and high mass stars at various radii, and the
consequence of this loss for the evolution of the present day mass
function (PDMF) away from the initial mass function (IMF).
Understanding the form of the IMF of the current globulars is relevant
for theories of star formation and globular cluster formation
\cite{fr77}, and the related issue of how the Galaxy was assembled.
The mass function of a stellar system $dN/dM$ is defined as the number
of stars $dN$ in the mass interval $(M,M+dM)$.  A common functional
form chosen to approximate this distribution is
\begin{equation}
	{dN \over dM} \propto M^{-(x+1)}
\label{mf}
\end{equation}
where $x$ is called the {\it mass function index}.  The PDMF $x$ of a
cluster has been found to be related to the Galactocentric radius
$R_{\rm GC}$ and height above the Galactic plane $Z$ of the cluster
(Capaccioli, Ortolani \& Piotto 1991).  
Using simple semi-analytic models Capaccioli, Piotto
\& Stiavelli (1993) showed that these correlations could be reproduced
by evolving a population of globular clusters which all have the same
initial IMF in the tidal field of the Milky Way.  Similarly, in a
comparison of several globular clusters Piotto, Cool \& King (1997)
found the luminosity function of NGC 6397 to be much flatter than M15,
M30 and M92, and concluded that this could be due to the extreme
dynamical evolution implied by its orbit calculated from its 
proper motion 
\cite{d96}.  In addition McClure et al. (1986) reported a dependence
of the PDMF on metallicity, and Djorgovski, Piotto \& Capaccioli
(1993) successfully incorporated this together with the trends in
$R_{\rm GC}$ and $Z$ into a trivariate analysis.  In contrast, in a
recent study of Pal 1, Rosenberg et al. (1997) found its mass function
to be inconsistent with the trends in $R_{\rm GC}$, $Z$, and
metallicity and concluded that this could either be due to evaporation
of low mass stars, or an intrinsically different IMF.

One step towards solving the puzzle of just how far the PDMF of a
cluster differs from its IMF is determining how to interpret the
signatures of tidal effects in observations of globular clusters and
other satellites.  In the past, the limiting radii of Galactic
satellites (which are assumed to correspond to the tidal radii $r_{\rm
tide}$ imposed by the Milky Way) have been interpreted using
King's (1962) tidal radius formula
\begin{equation}
	r_{\rm tide}=R_{\rm GC} 
        \left({m_{\rm sat} \over 3 M_{\rm GC}}\right)^{1\over 3},
\label{rtide}
\end{equation}
where $R_{\rm GC}$ is the distance of the satellite from the Galactic
centre, $m_{\rm sat}$ is the mass of the satellite, and $M_{\rm GC}$
is the mass of the Galaxy enclosed within $R_{\rm GC}$.  For the
observed $r_{\rm tide}$ this formula has been used to: (i) find
$m_{\rm sat}$ from the current $R_{\rm GC}$ \cite{fl83}; (ii)
estimate the pericentre of the satellite's orbit given the value for
$m_{\rm sat}$ implied by its internal dynamics (Oh, Lin \& Aarseth 1995; 
Irwin \&
Hatzidimitriou 1995 -- hereafter IH85); and (iii) compare the
Galaxy's tidal field with the satellite's internal field \cite{ih95}.
Such analyses have been complicated by the detection of
`extra-tidal' stars around both dwarf spheroidal galaxies
(IH95; Kuhn, Smith \& Hawley 1996) 
and globular clusters (Grillmair et al.
1995 -- hereafter G95; for a summary of current observations
of both Galactic and extra-galactic globular clusters, see
Grillmair 1998), and these morphological features have been shown to be
consistent with those seen in simulations of tidally disrupting
systems \cite{ola95,g95}.

Hut \& Djorgovski (1992) 
also estimated destruction rates of globular clusters by
finding the extent to which their distribution of half-mass
relaxation times deviates from a power law and interpreting this
difference in terms of evolution of the system.

An alternative to these very direct interpretations of observations is
to build models that include some representation of all the expected
dynamical effects on globular clusters and integrate the cluster's
evolution over the lifetime of the Galaxy \cite{aho88,ch96,go97,mw97,v97}.  
The success of
these semi-analytic methods (as with those in the previous paragraphs)
rests upon the representation of the physics involved, and there is
some disagreement about the level of sophistication required for
accurate predictions  -- one example is the theory of tidal shocking
which has received renewed attention \cite{w94,ko95,jhw98,gho98}.
However, all these studies predict that
the Milky Way's globular cluster system is currently undergoing
significant evolution.

In this paper, we return to the earlier philosophy of trying to
understand how much we can learn directly from observations of tidal
signatures.  We approach the problem by using numerical simulations to
assess how successfully such features can be interpreted.  Our study
differs from earlier ones in several ways: (i) we include a mass
spectrum, and begin with mass segregated models; (ii) there is a
one-one representation of stars in e.g.  globular clusters; and (iii)
the tidal field of a full Milky Way model is included rather than
idealised as giving rise to a spherical tidal boundary or being
represented by a one-component potential.  In our analysis, we
explicitly distinguish between the bound and unbound stars in a
satellite, which allows us to identify the characteristics of the
extra-tidal population.  We use this analysis to develop and test
methods for quantifying the mass loss rate and evolution of the mass
function of clusters and satellites directly from current
observations.

We present the simulation and analysis methods used throughout the
paper in \S 2.  The simulations are used to provide an overview of the
characteristics of evolution in a tidal field in \S 3.  We analyse our
models from the viewpoint of an observer \S 4.  In \S 5, the
interpretations developed in \S 4 are applied to observations and used
to measure the rate of destruction of the Milky Way's satellite
system.  We summarise our results in \S 6.

%%\clearpage
\section{METHODS}

\subsection{General approach}

In our calculations, the satellite is represented as a discrete set of
particles of different masses, one for each star expected in a cluster
of the chosen mass and mass function.  The particular choice of
cluster models in this paper follows the parameters adopted by
Chernoff \&\ Weinberg (1990) (see also Sigurdsson \& Phinney 1995).  
The initial
particle masses, positions and velocities are taken from an explicit,
maximal $N$ realisation of an isotropic Michie--King model of a
cluster \cite{k62,dc76,gg79}, as described in \S
2.2.  The cluster's evolution along an orbit in a three component
rigid model of the Galaxy (described in \S 2.3) is followed using a
self-consistent field (SCF) code to calculate the mutual interactions
of stars in the cluster (see \S 2.4.1).  The SCF approach cannot
follow evolution due to close encounters of individual stars in the
system, so we repeat one case with the effects of two-body encounters
included via diffusion coefficients to illustrate their importance
(see \S 2.4.2).

This study is intended to isolate the response of a cluster to tidal
effects with the dual purposes of identifying observable signatures of
tidal interactions and providing a set of control models to compare to
future studies which will include more of the physics that might
influence the cluster's evolution.  In particular, we make the
following simplifications:
\begin{enumerate}
	\item The distribution of particle masses is chosen to
represent an evolved stellar population with a turnoff at $\sim 0.8
M_{\odot}$.  We assume that continued stellar evolution is slow
compared to the dynamical timescales involved.  This assumption will
be relaxed in subsequent papers, with explicit stellar evolution done
in tandem with the dynamical evolution.
	\item The initial cluster models are truncated at some finite
radius, $r_T$, corresponding to the `tidal radius' of the King
model.  However, $r_T$ is not set to the limiting radius expected
along each orbit, as a key purpose of the study is to investigate
observable effects of tidal shocks and stripping of a cluster that is
not in exact equilibrium with the tidal field.  The chosen models and
orbits cover a range of interaction strengths -- from those that only
survive a few orbits to those that could last for the lifetime of the
Galaxy.  The scenario implicit in these initial conditions is that a
cluster is formed very compact, with most of its mass well inside its
actual tidal radius. Mass loss during stellar evolution subsequent to
the cluster's formation leads to expansion of the cluster, until the
tidal radius is reached. Here we assume that there is negligible loss
of stars due to tidal effects until late in the cluster's evolution,
and that we can start with a relaxed, mass segregated cluster with an
exact King model profile, and follow subsequent mass loss due to tidal
effects.  Clearly this is an approximation to the real physics, but in
order to isolate the tidal effects we are exploring, the initial
conditions of the cluster must be close to equilibrium or internal
dynamical evolution will instead dominate.
	\item As noted above, the effects of two body encounters on the
internal dynamical evolution have not been included in the majority of
our models.
\end{enumerate}
In this paper we restrict detailed discussions to effects that are
insensitive to these simplifications.  For example, in \S 4 we use our
models to test how a cluster's extra-tidal population can be used to
determine its mass loss rate -- we expect this to be valid because
although the mass loss rate itself will depend on the internal
dynamics of the system (which we do not model exactly), the
characteristics of the extra-tidal population are determined only by
the tidal potential (which we do model exactly).  Conversely, we do
not attempt to arrive at quantitative conclusions in our discussion of
the evolution of clusters along different orbits (\S 3) since these
would be influenced by all three of the above simplifications.

\subsection{Initial cluster models}

\begin{figure}
\begin{center}
\epsfxsize=8cm \epsfbox{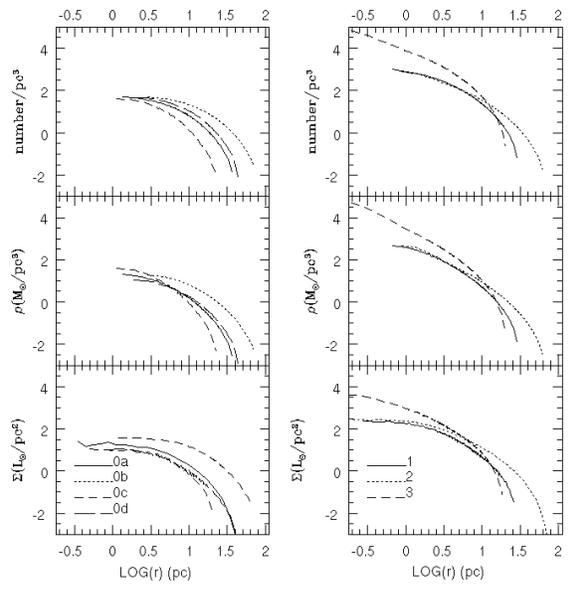}
\caption{ Initial number density (top panels), mass density (middle panels),
and surface density (bottom panels) for each of the initial models.}
\label{profifig}
\end{center}
\end{figure}

\begin{figure}
\begin{center}
\epsfxsize=8cm
\epsfbox{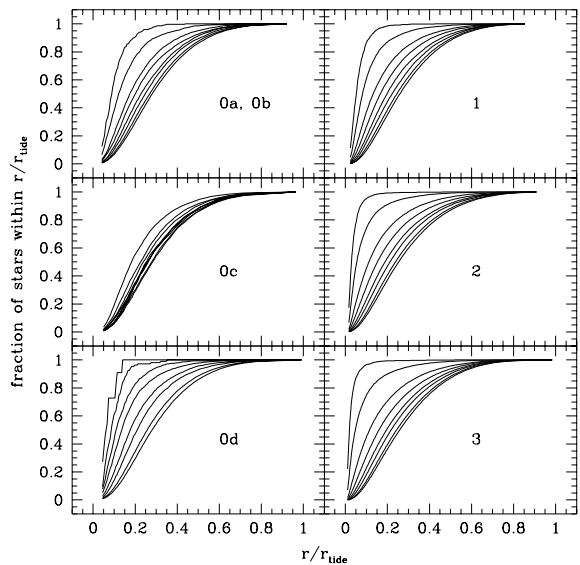}
\caption{Cumulative distribution of stars for each mass group in the
labelled models.
The highest/lowest curves correspond to the largest/smallest mass group
(i.e. bin number $\alpha$
increasing upwards).
The mass range encompassed by a curve representing mass bin $\alpha$
can be found by multiplying
the mass of the model (column 2 of Table \ref{modstab})
by the mass fraction $dm_\alpha$ in bin $\alpha$ for the
appropriate mass function index $x$ (see Table \ref{startab}).
\label{cummfig}}
\end{center}
\end{figure}

\begin{table*}
\begin{minipage}{16cm}

\caption{ Description of Models.}
\label{modstab}
\begin{tabular}{|cccccccccc|}
\hline
Model & mass & $r_{\rm half}$ & $N$ & $W_{0}$ & \# density & $x$ & $T_{\rm dyn}$ 
& $T_{\rm relax}$  & orbits \\ 
      & $10^5 M_{\odot}$ & pc &&   & \#$/pc^3$  && $10^6$ years & 
$10^9$ years &\\   
\hline
\hline
0a    & 0.34284 & 10.6 & 142365  & 4 & 50                 & 1.35 & 6.19 & 2.68 
& p1-3,d1-2 \\
0b    & 2.73969 & 21.2 & 1136640 & 4 & 50                 & 1.35 & 6.19 & 17.66
& p1 \\
0c    & 0.20614 & 6.24& 29904   & 4 & 50                 & 0.00 & 3.61 & 0.383 
& p1 \\
0d    & 0.43807 & 13.4 & 269366  & 4 & 50                 & 2.50 & 7.74 & 6.00 
& p1 \\
1     & 1.22626 & 7.36 & 508669  & 6 & 1000               & 1.35 & 1.89 & 2.62 
& p1,p2,d1 \\
2     & 2.86636 & 14.1 & 1186585 & 9 & 1000               & 1.35 & 3.28 & 9.92 
& p1 \\
3     & 3.30330 & 4.62 & 1368230 & 12 & 1.0 $\times 10^5$ & 1.35 & 0.573 & 1.98
& p1,p2 \\ 
\hline
\end{tabular}

\medskip
Columns: (1) Model number; (2) mass; (3) half mass radius;
(4) number of stars; (5) King model; (6) central number density;
(7) mass function index; (8) internal dynamical time (see equation
[\ref{tdyn}]);
(9) half-mass relaxation time; (10) orbits simulated.
\end{minipage}
\end{table*}
Four representative cluster models (labelled 0-3) are considered, with
properties summarised in {Table \ref{modstab}}.  Model 0a was repeated
with the same number density profile, but larger mass (Model 0b), and
with different mass function indices (Models 0c and 0d).  {Figure
\ref{profifig}} shows the number density and mass density as a
function of spherical radius, and surface brightness as a function of
projected radius for each initial distribution.  Masses are converted
to visual magnitudes and luminosities in this and subsequent plots
using Bergbusch \& VandenBerg's (1992) isochrone 
for a 14 Gyr, [Fe/H]=-1.66 cluster.  {
Figure \ref{cummfig}} illustrates the level of mass-segregation by
showing the cumulative fraction of stars within a given radius for
each mass group in the models.

The cluster is generated as a Monte Carlo realisation of the
Michie--King distribution function, using 8 discrete mass groups and a
truncated, evolved power law initial mass function as described in
Sigurdsson \&\ Phinney (1995).  The structure of each cluster is
determined from a set of input parameters: the depth of the 
central potential, $W_0$,
the central number density, $n_0$, and the mean central dispersion,
$\sigma $.  Some of the observable structure parameters, such as the
concentration, core radius and light profile,
are derived quantities and depend on
the mass function and mass--luminosity relation chosen.

The particles representing the stars in mass group $\alpha (= 1\ldots
8)$, of mass $m_{\alpha}$, are assumed to have some initial
distribution function, $g_{\alpha} (\varepsilon )$, where
$\varepsilon=-\Phi-{1\over 2}{\bf v}^2$ is the energy and ${\bf v}$ is
the velocity of the particle (both in the cluster centre--of--mass
frame), and $\Phi$ is the cluster potential. The distribution function
is then given by
\begin{equation}
g_{\alpha} (\varepsilon ) = {{n_{0_{\alpha}}}\over 
{(2\pi \sigma_{\alpha})^{3/2}}}  \Bigl [ e^{{{
\varepsilon }/{\sigma_{\alpha} }}} - 1 \Bigr ],
\end{equation}
where $\sigma_{\alpha}$ is the core dispersion of mass group $\alpha $
and the $n_{0_{\alpha}}$ are the normalised densities of each mass
group.  The $n_{0_{\alpha}}$ are solved iteratively given the cluster
structure parameters, according to the scheme described in Sigurdsson
\&\ Phinney (1995).

\begin{table*}
\begin{minipage}{16cm}
\caption{ Adopted Mass Functions with Index $x$}
\label{startab}
\begin{tabular}{|c|ccc|ccc|ccc|}
\hline
$\alpha$        && $x=0.0$              &&& $x=1.35$            &&& $x=2.50$ &
\\
& $m_{\alpha}$ & $dm_{\alpha}$ & $f_{\alpha}$ & $m_{\alpha}$ & $dm_{\alpha}$ & 
$f_{\alpha}$ & $m_{\alpha}$ & $dm_{\alpha}$ & $f_{\alpha}$ \\
\hline
\hline
  1  & 0.12629  & 0.01847 & 1.00000 & 0.12346  & 0.23393  & 1.00000 & 0.12114  & 
0.50252 & 1.00000 \\
  2  & 0.22461  & 0.04963 & 1.00000 & 0.21330  & 0.29045  & 1.00000 & 0.20449  & 
0.33330 & 1.00000 \\
  3  & 0.34806  & 0.02601 & 1.00000 & 0.34600  & 0.08360  & 1.00000 & 0.34427  & 
0.05484 & 1.00000 \\
  4  & 0.44265  & 0.03596 & 1.00000 & 0.43956  & 0.08355  & 1.00000 & 0.43695  & 
0.04163 & 1.00000 \\
  5  & 0.57392  & 0.12309 & 0.34155 & 0.56679  & 0.13430  & 0.52623 & 0.56126  & 
0.03946 & 0.67655 \\
  6  & 0.71009  & 0.16844 & 0.32948 & 0.70416  & 0.11552  & 0.58798 & 0.70078  & 
0.02525 & 0.77595 \\
  7  & 0.99995  & 0.22745 & 0.00000 & 0.96585  & 0.04281  & 0.00000 & 0.93946  & 
0.00264 & 0.00000 \\
  8  & 1.38462  & 0.35095 & 0.00000 & 1.36336  & 0.01583  & 0.00000 & 1.34231  & 
0.00036 & 0.00000 \\
\hline
\end{tabular}

\medskip
Columns: (1) -- mass group; (2), (5) and (8) -- average mass $m_{\alpha}$; 
(3), (6) and (9) -- mass fraction 
$dm_{\alpha}$;
(4), (7) and (10) -- fraction of luminous stars assigned to bin $\alpha$.
\end{minipage}
\end{table*}
For each mass function index $x$ considered in this paper, Table
\ref{startab} gives the average mass $m_{\alpha}$ of stars and total
mass fraction $dm_{\alpha}$ assigned to bin $\alpha$, along with the
fraction $f_{\rm \alpha}$ of stars in each bin that are luminous.
Stars that have evolved beyond the main--sequence turnoff are assigned
to remnant stellar classes. In particular, stars with initial masses
between the turnoff and $4.7 M_{\odot}$ are assumed to have evolved to
white dwarfs, and stars with initial masses between $8$ and $15
M_{\odot}$ are assumed to have formed $1.4 M_{\odot}$ neutron stars.
This follows the prescription in Chernoff \& Weinberg (1990). Future
work will consider other prescriptions for assigning remnant masses as
a function of initial stellar mass.  The presence of a substantial
population of intermediate mass white dwarfs changes the present day
mass function from the simple power law of the initial mass function.
In particular, there is a pronounced `bump', an excess of stars with
masses $\sim 0.5 M_{\odot}$ in the distribution function, which
complicates interpretation of the local luminosity function in terms
of the mass function, for intermediate mass stars.

Each cluster is generated by a simple acceptance--rejection algorithm,
selecting particle masses, positions and velocities from the
distribution function.
The masses are assigned in discrete mass bins, with the value
of the mass in each bin being the mean mass of that interval, weighted
by the evolved mass function. That is, the mean mass in each bin
allows for the presence of stellar remnants in that bin.
Future models may incorporate a
continuous distribution of stellar masses.
The positions are drawn
from a radial grid (typically of about 200 points, spaced to sample
the density profile efficiently). A particle selected to be at radius
$r_i$ is assigned to some radius $r_i \pm \epsilon$ where $\epsilon$
is a uniform random variable on the grid space interval.  The initial
distribution is thus a set of thin stepped radial shells, with local
density that deviates slightly from the true density profile.  Phase
mixing ensures that the model settles down to a smooth representation
of the cluster on a dynamical timescale.  The initial model has a
virial ratio $\sim 1 \pm 1/\sqrt{N}$, and is stationary and stable.
Cluster models were run in isolation using the SCF code (see \S 2.4.1
and Hernquist \& Ostriker 1992): 
the initial mass segregation is robust and there is
no spurious evolution.  The mass segregation corresponds to thermal
equilibrium for the cluster structure parameters (central potential and
initial mass function) chosen. This is the equilibrium state expected
for the internal distribution of masses within the cluster once the
time scale for mass loss due to stellar evolution becomes longer than
the core relaxation time, and provided that tidal effects have not yet
had a strong effect on the cluster.  There is also some evidence for
primordial mass segregation in young clusters \cite{f98,e98}, 
possibly due to bias in the formation of massive stars
towards high density regions, or due to rapid relaxation in the young
cluster. Mass segregation continues on a relaxation time scale, and
would lead to core collapse in the absence of other effects.

\subsection{Cluster orbits in the Galactic potential}

A three-component model is used to represent the Galactic potential:
$\Psi=\Psi_{\rm disk}+\Psi_{\rm spher}+\Psi_{\rm halo}$, in which the
disk is represented by a Miyamoto-Nagai potential \cite{mn75}, the
spheroid by a Hernquist potential \cite{h90}, and the halo by a
logarithmic potential:
\begin{equation}
        \Psi_{disk}=-{GM_{\rm disk} \over
                 \sqrt{R^{2}+(a+\sqrt{z^{2}+b^{2}})^{2}}},
\label{disk}
\end{equation}
\begin{equation}
        \Psi_{\rm spher}=-{GM_{\rm spher} \over r+c},
\label{bulge}
\end{equation}
\begin{equation}
        \Psi_{\rm halo}=v_{\rm halo}^2 \ln (r^{2}+d^{2}).
\label{halo}
\end{equation}
Here, $M_{\rm disk}=1.0 \times 10^{11}, M_{\rm spher}=3.4 \times
10^{10}, v_{\rm halo}= 128, a=6.5, b=0.26, c=0.7$, and $d=12.0$, where
masses are in $M_{\odot}$, velocities are in km s$^{-1}$ and lengths are in
kpc. This choice of parameters provides a nearly flat rotation curve
between 1 and 30 kpc and a disk scale height of $0.2 $ kpc.  The
radial dependence of the z epicyclic frequency ($\kappa_z$) in the
disk between radii at 3 and 20 kpc is similar to that of an
exponential disk with a 4 kpc scale length.

\begin{figure}
\begin{center}
\epsfxsize=8cm
\epsfbox{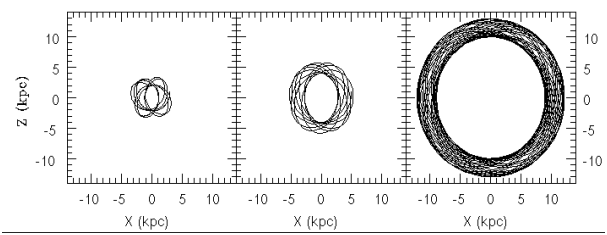}
\caption{Locus of polar orbits p1 (left hand panel), p2 (middle panel) and
p3 (right hand panel). The Galactic plane is at $Z=0$.
\label{orbpfig}}
\end{center}
\end{figure}

\begin{figure}
\begin{center}
\epsfxsize=8cm
\epsfbox{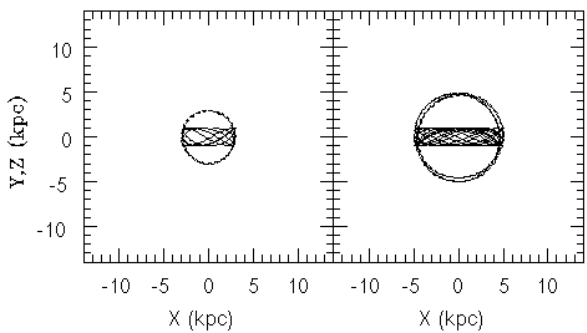}
\caption{Locus of disk orbits d1 (left hand panel) and d2 (right hand panel).
The solid line is for motion perpendicular to the Galactic plane (i.e.
$Z-X$ motion) and the dotted line is for motion in the Galactic plane 
(i.e. $Y-X$ motion).
\label{orbdfig}}
\end{center}
\end{figure}

\begin{table*}
\begin{minipage}{16cm}
\caption{ Description of Orbits.}
\label{orbstab}
\begin{tabular}{|cccccccc|}
\hline
Orbit & $r_{\rm peri}$ & $r_{\rm apo}$ & $T_{\rm orb}$  & $A_{\rm disk}$ 
& $T_{\rm disk}$ & $A_{\rm bulge}$ & $T_{\rm bulge}$ \\ 
      & kpc        & kpc       & $10^7 years$ & (km s$^{-1}$)/kpc/Myear 
& $10^6 years$ & (km s$^{-1}$)/kpc/Myear & $10^6 years$ \\
\hline
\hline
p1    & 1.1        & 3.5       & 8.0            & 30       & 1.0 & 20-30 & 3.
\\
p2    & 3.0        & 5.8       & 14.5           & 20       & 1.0 & 2-3   & 10.
\\
p3    & 9.5        & 12.5      & 32.0           & 5-10     & 1.0 & 0.5   & 20.
\\
d1    & 2.9        & 3.15      & 8.5            & 30       & 2.0 & 3-4   & 12.
\\
d2    & 4.6        & 5.4       & 14.5            & 30       & 2.2 & 1     & 
20.\\
\hline
\end{tabular}

\medskip
Columns: (1) orbit; (2) closest approach; (3) apocentre;
(4) azimuthal orbital time period; (5) defined in equation (\ref{a});
(6) timescale for disk passage (see equation [\ref{tdisk}]);
(7) and (8) as columns (5) and (6) but for bulge passages.
\end{minipage}
\end{table*}
To contrast the evolution of the models in different tidal fields we
choose orbits roughly 3kpc, 5kpc and 10kpc from the Galactic centre,
three with polar orientations (p1, p2 and p3) and two near the
Galactic disk (d1 and d2).  Figures \ref{orbpfig} and \ref{orbdfig}
show the paths of these orbits and Table \ref{orbstab} gives their
peri- and apo-galacticon and radial time periods.

Hereafter we refer to the simulation representing the evolution of
cluster Model $m$ along orbit $o$ as Model $(m,o)$.
 
\subsection{Integration methods}

Individual particle trajectories are integrated using a leapfrog
integration scheme, with accelerations calculated from
\begin{equation} 
     \label{accel} {\bf \ddot r} = \nabla 
     (\Psi ({\bf r})+\Phi({\bf r})) + {\bf a_{dyf}} + {\bf a_{kick}},
\end{equation}
where the Milky Way is represented by the rigid potential $\Psi$ given
in \S 2.2, and the cluster's internal potential $\Phi$ is calculated
from the particles' masses and positions using the SCF code described
in \S 2.4.1.  The effect of accelerations due to two-body effects
(i.e. terms ${\bf a_{dyf}} + {\bf a_{kick}}$) were included in only
one simulation using the method described in \S 2.4.2.  Dynamical
friction of the cluster's orbit was ignored.  In all cases, the time
step was smoothly varied along the orbit between $T_{\rm disk}/100$
and $T_{\rm dyn}/100$ to ensure that energy was conserved to better
than 1 percent of the initial internal potential energy of the cluster.
Here
\begin{equation}
	T_{\rm disk}=b/v_z
\label{tdisk} 
\end{equation}
is the timescale for a disk passage, where $b$ is the vertical scale
of the Galactic disk (see equation [\ref{disk}]), and $v_z$ is the
z-component of the satellite's velocity as it crosses the disk plane
at $Z=0$, and
\begin{equation}
\label{tdyn}
	T_{\rm dyn}=\pi \sqrt{r_{\rm half}^3 \over 2 G m_{\rm sat}},
\end{equation}
is the internal dynamical timescale for a system of total mass $m_{\rm sat}$ 
and
half-mass radius $r_{\rm half}$
\cite{bt}.  The timescales for the orbits and
models are given in Tables \ref{orbstab} and \ref{modstab} respectively.

\subsubsection{Self-consistent field code}

The SCF code uses a bi-orthogonal basis function expansion to
calculate the internal potential of the cluster from the individual
particle positions and masses \cite{ho92}.  The expansion is
sensitive to large-scale fluctuations in the cluster's potential but
smooths local potential fluctuations arising from the discrete
particles.  Hence the SCF approach will underestimate relaxation when
(as in our case) the number of particles in the simulation is
equivalent to the number of stars in the system represented.  The
computations using this scheme (i.e. not including two-body
encounters) were performed at the National Center for Supercomputing
Applications (NCSA) with a version of the SCF code which had been
parallelised to run on the Connection Machine 5 \cite{her95}.  On
average, each step required $4\times
10^{-4}$cpusec/particle/processor.

\subsubsection{Two-body relaxation calculation}

In order to explore the consequences of continuing mass segregation
and two--body relaxation on the structure of the cluster, we
incorporated a Fokker--Planck diffusion scheme into a version of the
SCF code parallelised to run on the T3E \cite{ss97} at the
Pittsburgh Supercomputing Center.  The
results of Model (0a,p3), run with and without diffusion are compared
in in \S 3.3.

The Fokker--Planck scheme calculates the first and second order
velocity diffusion coefficients, $D(\Delta v_i), D(\Delta v_i\Delta
v_j)$ \cite{bt}.  The diffusion is directly incorporated into the
explicit time evolution of the particles by including the last two
terms in equation (\ref{accel}), where ${\bf a_{dyf}} $ is the
dynamical friction experienced by the star,
and ${\bf a_{kick}}$ is the effective acceleration due to
scattering by individual stars in the cluster (see Sigurdsson \&
Phinney 1995 for discussion).

To calculate ${\bf a_{dyf}} $ and ${\bf a_{kick}}$, we choose an
orthonormal basis local to each particle defined by the particle's
velocity ${\bf v}$ and position ${\bf x}$ in the cluster frame.  This
gives us three independent diffusion coefficients, $D(\Delta
v_{\parallel })$, $D(\Delta v^2_{\parallel})$, and $D(\Delta
v^2_{\perp})$.  To calculate those we need the local density, which is
obtained directly from the SCF expansion \cite{ho92}, and a local
velocity distribution. We approximate the velocity distribution as a
Gaussian, and calculate it every $N_v \sim 100$ integration steps, by
sampling the velocity distribution in radial bins, calculating the
local dispersion, and using spline interpolation to obtain the
dispersion as a function of radius.  Provided the dynamical evolution
of the cluster is slow compared to the dynamical time scale, the
sparse updating of the dispersion profile is adequate, and provided
the cluster is close to being relaxed, the approximation of taking the
velocity distribution as an isotropic Gaussian is acceptable. The
diffusion approximation incorporates a Coulomb logarithm term whose
magnitude is uncertain to a magnitude comparable to other errors
incurred by the approximations we make.  Near the truncation radius,
the interpolation of the dispersion must be positive definite or
negative dispersion may be inferred.  Outside the truncation radius
the diffusion coefficients may be set to zero, or the local dispersion
approximated by the Keplerian velocity due to the total enclosed mass.

Given the diffusion coefficients, then ${\bf a_{dyf}} = D(\Delta
v_{\parallel })$ and we model ${\bf a_{kick}}$ by random fluctuations
in velocity, $\Delta {\bf v}$, where $ {\bf a_{kick}} = {{\Delta {\bf
v}}\over {\Delta t}}$, with
\begin{eqnarray}
\Delta v^2_{\parallel } &= \varsigma_i ^2 (D(\Delta v^2_{\parallel }
) \Delta t ) \\
\Delta v^2_{\perp } &= \varsigma_i ^2 (D(\Delta v^2_{\perp }) \Delta t ),
\end{eqnarray}
where $\varsigma_i $ is a random number with zero mean and unit
standard deviation, chosen here from a normal distribution.

This choice of diffusion coefficients provides a fast and surprisingly
good approximation to two--body relaxation in clusters, and allows
a fair representation of the effects of two--body relaxation on the
structure of the cluster in the presence of other perturbing dynamical
processes. For isolated clusters, this scheme provides a realisation
of evolution towards core collapse on relaxation time scales, and can
follow the evolution of the cluster over several orders of magnitude
in central density. Ultimately though, with a finite number of
expansion terms, the SCF code fails to resolve the resulting density
cusp and core collapse to infinite density is not observed. Details of
the scheme will be discussed in another paper.

\subsection{Analysis methods}

\subsubsection{Isolating the bound population of the cluster}

In parts of this study, identical analyses are performed on the
particle data from the simulations both with and without the unbound
particles, and the results are compared.  Much of the discussion of
the interpretation of observations rests upon the identification of
stars that are cluster members.  We define members of the cluster to
be the maximum set of mutually bound particles.  This set is found
iteratively, starting from the set of all particles, by: (i)
calculating the internal potential field of the particles in the set
considered; (ii) finding the velocity of each particle with respect to
the velocity of the minimum of this potential field; (iii) defining
each particle's internal kinetic energy to be the kinetic energy of
this relative motion; (iv) labeling those particles whose internal
kinetic energy is greater than their internal potential energy as
`unbound', removing them from the set considered and returning to
step (i).  Steps (i)-(iv) are repeated until no new particles in an
iteration are labelled `unbound'.  Using this method we find that the
instantaneous mass bound to the cluster is a monotonically decreasing
function of time, and the number of stars that later become bound
again to the cluster once they have first been classified as unbound
is negligible.

\subsubsection{Extrapolating mass functions}

Each particle in our simulations is assigned the average mass
$m_\alpha$ of the mass bin $\alpha$ that it occupies.  The first four
mass bins contain only luminous matter, the next two contain a
fraction $f_\alpha$ of luminous stars and the two most massive bins
contain only stellar remnants (see Table \ref{startab}).  Hence, we can
trivially find the mass function of observable stars in any sample at
these average points by calculating
\begin{equation}
	\left({dN \over dm}\right)_\alpha=
	{f_\alpha N_\alpha \over \Delta m_\alpha},
\label{dndm}
\end{equation}
where $\Delta m_\alpha$ is the width of the mass bin and $N_\alpha$ is
the number of stars in the bin.  Since mass functions are typically
calculated from stars near the turnoff in a globular cluster, we
estimate $x$ using only the two heaviest luminous mass bins to be
\begin{equation}
	x={LOG(N_6 f_6 \Delta m_5)-LOG(N_5 f_5 \Delta m_6) \over
	   LOG(m_6)-LOG(m_5)}-1.
\label{x}
\end{equation}

%%\clearpage
\section{GENERAL EVOLUTION IN A TIDAL FIELD}

\subsection{Cluster heating}

We expect mass loss to occur predominantly where the tidal field of
the Milky Way is strongest -- i.e. at pericentric points along the
orbit and during disk passages.  In column 5 of Table \ref{orbstab} we
give values for
\begin{equation}
	A_{\rm disk}=d^2 \Phi_{\rm disk}/dz^2.
\label{a}
\end{equation}
For a globular cluster whose physical scale is given by its half mass
radius $r_{\rm half}$, this can be used to estimate the specific
energy change due to the disk passage in the impulsive regime,
\begin{equation}
\Delta E_{\rm imp} = {1\over 2} \left(A_{\rm disk}
T_{\rm disk}r_{\rm half}\right)^2,
\label{eimp}
\end{equation} 
\cite{s58} which we shall subsequently refer to as the {\it
shock strength}, where $T_{\rm disk}$ is the timescale for the disk
passage (see equation [\ref{tdisk}]).  We expect this estimate to be
appropriate when $T_{\rm dyn} \gg T_{\rm disk}$, and can otherwise use
it as an upper limit on the average energy change.  Columns 7 and 8 in
Table \ref{orbstab} give similar estimates for the importance of bulge 
shocking for these orbits (with $A_{\rm bulge}=d^2\Phi_{\rm bulge}/dR^2$),
which is clearly competitive with disk shocking for
orbit p1 (see Aguilar, Hut \& Ostriker 1988
for a discussion of this effect).  In our
chosen Galactic potential, all these orbits lie within the core radius
$d=12$kpc of the halo component (see equation [\ref{halo}]), so its
tidal field is roughly constant along the orbit and halo shocking is
unimportant (in contrast, this is likely to be an important factor in
the evolution of the Sagittarius dwarf spheroidal galaxy -- see
Johnston, Spergel \& Hernquist 1995).

\begin{figure}
\begin{center}
\epsfxsize=8cm
\epsfbox{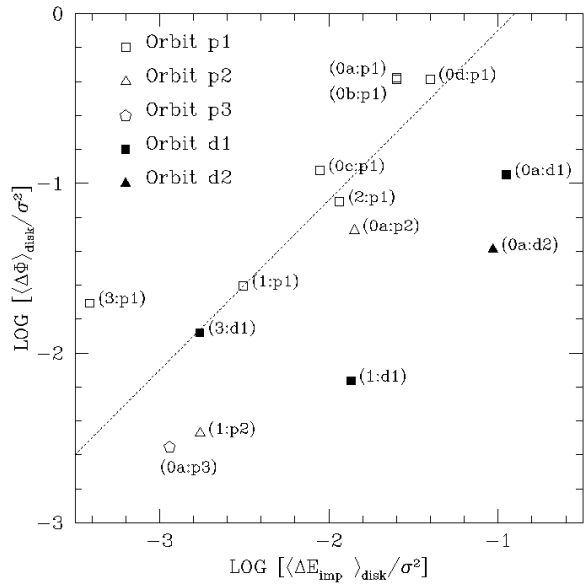}
\caption{Average potential change per disk shock for each model as
a function of shock strength, in units of the clusters' internal velocity 
dispersion $\sigma$. Each point is labelled with the model and orbit.
The dotted line indicates the trend expected when the
impulse approximation is valid.
\label{dphifig}}
\end{center}
\end{figure}
To test the accuracy of the impulse approximation, we calculated the
change in the cluster's internal potential energy per unit mass
($\Delta \Phi$) and the impulsive estimate for this change ($\Delta
E_{\rm imp}$ -- from equations [\ref{tdisk}], [\ref{a}] and
[\ref{eimp}]) for each of the $N_{\rm disk}$ disk passages in each
simulation.  The points in {Figure \ref{dphifig}} show the average
values of these quantities in the simulations, in units of the
cluster's velocity dispersion, $\sigma^2$.  The models in the bottom
left hand corner of the plot are expected to be the most resilient to
the Milky Way's influence.

There is a large spread in the response of the clusters' potential
energy for a given shock strength, $\Delta E_{\rm imp}$.  The clusters
on disk orbits (filled symbols) are shocked less noticeably than those
on polar orbits (open symbols), because the timescales for the disk
passages in the former case are longer, and the impulse approximation
will only be valid in the outer region of the cluster.  In general,
Figure \ref{dphifig} shows that while this crude estimate serves as a
useful guide to the importance of shock heating for a cluster in a
given orbit a much more detailed calculation of the interaction of the
internal and external dynamics is needed for an accurate assessment of
cluster evolution in general.  More rigorous analytic estimates for
tidal heating have been discussed extensively elsewhere in the
literature \cite{w94,ko95,go97,jhw98}.

\subsection{Differential mass loss}

\begin{figure}
\begin{center}
\epsfxsize=8cm \epsfbox{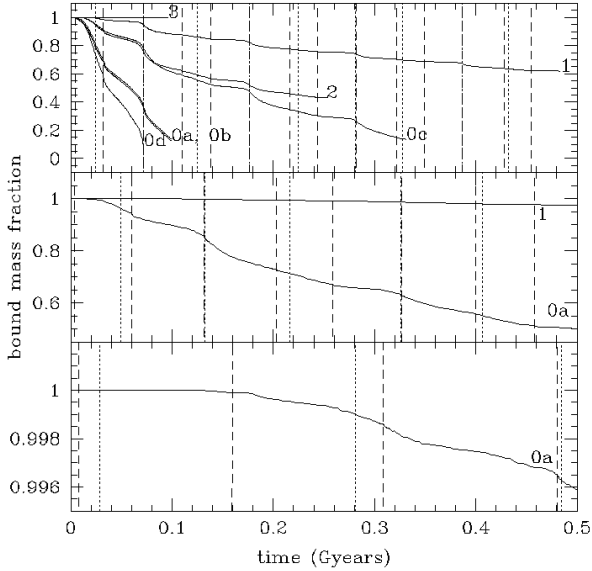}
\caption{Bound mass fraction as a function of time for the models
along the polar orbits p1 (top panel), p2 (middle panel) and 
p3 (lower panel). The dashed lines label the time of disk passages
and the dotted lines label the time of pericentric passages.
\label{masspfig}}
\end{center}
\end{figure}
\begin{figure}
\begin{center}
\epsfxsize=8cm \epsfbox{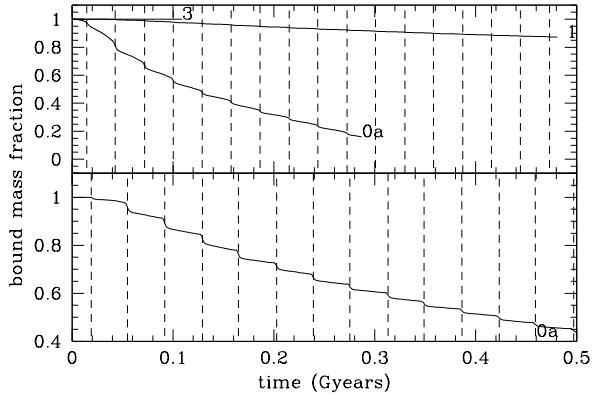}
\caption{Bound mass fraction as a function of time for the models
along the disk orbits d1 (top panel) and
d2 (bottom panel). The dashed lines label the time of disk passages.
\label{massdfig}}
\end{center}
\end{figure}
{Figures \ref{masspfig} and \ref{massdfig}} follow the bound mass
fraction remaining as a function of time for the first 0.5 Gyrs of
evolution along the polar and disk orbits.  On each of these figures
we show the time of disk passages as vertical dashed lines, and the
pericentric points along the orbits as vertical dotted lines.  As
noted in the previous section, the mass loss rate is greatest at these
locations along the orbit.  A comparison of these plots with Figure
\ref{dphifig} confirms the loose correlation between mass loss rate
and shock strength.

\begin{figure}
\begin{center}
\epsfxsize=8cm \epsfbox{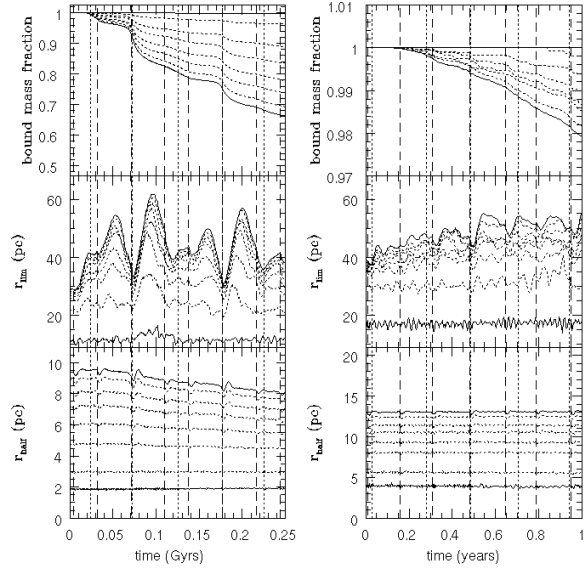}
\caption{Bound mass fraction, limiting and half mass radii for each
mass group as a function
of time for Model (1,p1) (left hand panels) and Model (0a,p3).
The vertical dashed and dotted lines give the time of 
disk and pericentric passages respectively.
The outer
solid lines in each panel are for the least and most massive stars
and the dotted lines are for the intermediate mass stars.
\label{evolfig}}
\end{center}
\end{figure}
{Figure \ref{evolfig}} compares the effect of the tidal field on each
individual mass group in Model (1,p1) (left hand panels) with those in
Model (0a,p3) (right hand panels).  The solid lines are for the
smallest and largest masses and the dotted lines are for the
intermediate ones.  The top panels illustrate differential mass loss
due to mass segregation, as the most weakly bound (and therefore lower
mass) stars are preferentially stripped. The lower two panels give a
measure of the response in the spatial distributions - $r_{\rm lim}$
is calculated as the average radius of the outermost 1 percent of bound
particles in each group, and $r_{\rm half}$ is calculated as the
radius containing half the mass of each group.  Again, the compressive
disk shocks (and subsequent expansion due to the energy input) are
clearly seen. The effect is more dramatic in the left hand panels
since this model is more susceptible to the influence of tides (as
demonstrated by its position in Figure \ref{dphifig}, and large mass
loss rate). Similar characteristics are seen in all the models.

\begin{figure}
\begin{center}
\epsfxsize=8cm \epsfbox{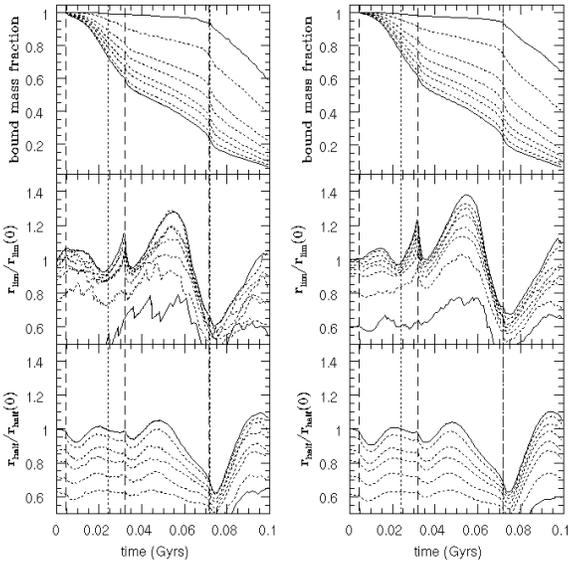}
\caption{As Figure \ref{evolfig} but for Models (0a,p1) and (0b,p1).
\label{m0abfig}}
\end{center}
\end{figure}
{Figure \ref{m0abfig}} repeats Figure \ref{evolfig} for Models (0a,p1)
(left hand panels) and (0b,p1) (right hand panels).  These models had
identical mass density profiles, but with different radial scales (and
hence total masses).  The comparison of the two demonstrates that the
general characteristics of the response for models with the same
density profiles subjected to the same tidal field are identical.

\begin{figure}
\begin{center}
\epsfxsize=8cm \epsfbox{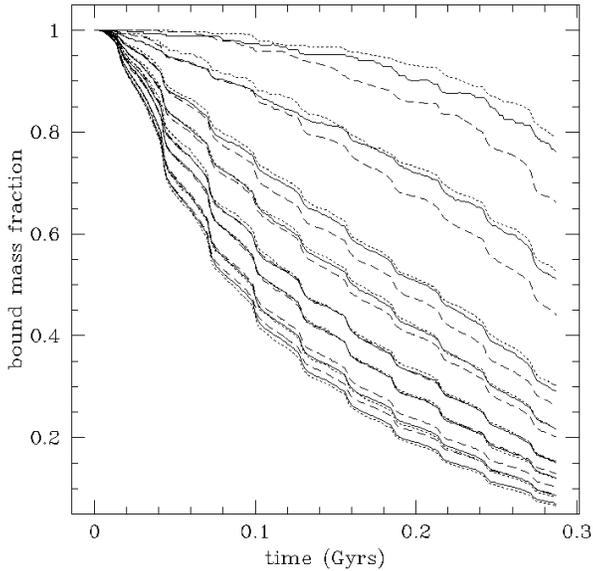}
\caption{The solid lines show the evolution of the bound mass fraction
for each mass group in Model (0a,d1).
The dashed lines show the fraction of each mass group within a  
spherical surface in the initial model
which contains the same total mass 
as is instantaneously bound to the satellite. 
The dotted lines show the fraction of each mass group within an
energy surface in each initial model
which contains the same total mass 
as is instantaneously bound to the satellite. 
\label{mevolfig}}
\end{center}
\end{figure}
\begin{figure}
\begin{center}
\epsfxsize=8cm \epsfbox{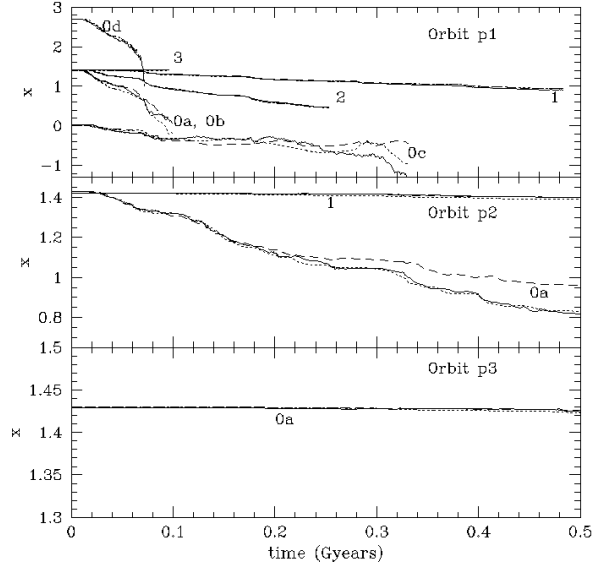}
\caption{The solid lines show the evolution of the mass function index
$x$ along each of the polar
orbits.
The dashed lines give $x$ for the
mass within a spherical surface in each initial model equivalent to
the bound mass remaining in the simulations at that time. 
The dotted lines show $x$ within an equivalent
energy surface in each initial model.
\label{xevolpfig}}
\end{center}
\end{figure}
\begin{figure}
\begin{center}
\epsfxsize=8cm \epsfbox{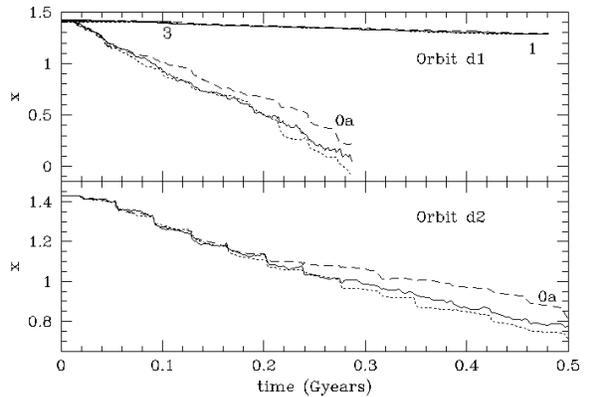}
\caption{As Figure \ref{xevolpfig} but for the disk orbits.
\label{xevoldfig}}
\end{center}
\end{figure}
Figures \ref{mevolfig} -- \ref{xevoldfig} illustrate our intuitive
understanding of tidal mass loss in a mass-segregated system.  As an
example, the solid lines in Figure \ref{mevolfig} show the bound mass
fraction for each mass group as a function of time for Model (0a,d1).
The dashed lines show the prediction if the initial model were simply
truncated in radius at the surface enclosing the same mass as is
instantaneously bound to the system in the simulation.  The dotted
lines show the prediction if the initial model is instead truncated in
binding energy.  As might be anticipated, initial binding energy
provides the more accurate of these simple guides to which stars are
lost first from the system.  This behaviour is seen in all our
simulations, as shown in {Figures \ref{xevolpfig} and
\ref{xevoldfig}}.  The solid lines show the global mass function index
$x$ as a function of time for each simulation, the dashed lines show
the result of estimating $x$ by truncating the initial model in radius
and the dotted lines show $x$ for truncation in energy.  Note in
general that the agreement is good, except when there has been
substantial evolution of the system (e.g Models (0a,p1), (0c,p1)).

In summary, the results in this section demonstrate that: (i) the
impulse approximation provides a rough guide to the mass loss rate
from a system; (ii) tidal mass loss along a given orbit is determined
by the density profile of the cluster; and (iii) tidal shocking of a
mass-segregated system leads to evolution of the mass function, which
can approximately be predicted for a given mass lost by simply
truncating the initial distribution in energy.  Clearly these general
trends fit in with theoretical interpretations of observed
correlations of cluster properties with $R_{\rm GC}$ and $Z$ 
\cite{cop91,cps93,dpc93}.  However, a quantitative
comparison of our simulations with observations would require more
detailed analytic models both of the tidal effects seen in the
simulations and of other physical processes not included (one example
is given in the next section) and is beyond the scope of the current
study.

\subsection{Competition of tides and relaxation}

\begin{figure}
\begin{center}
\epsfxsize=8cm \epsfbox{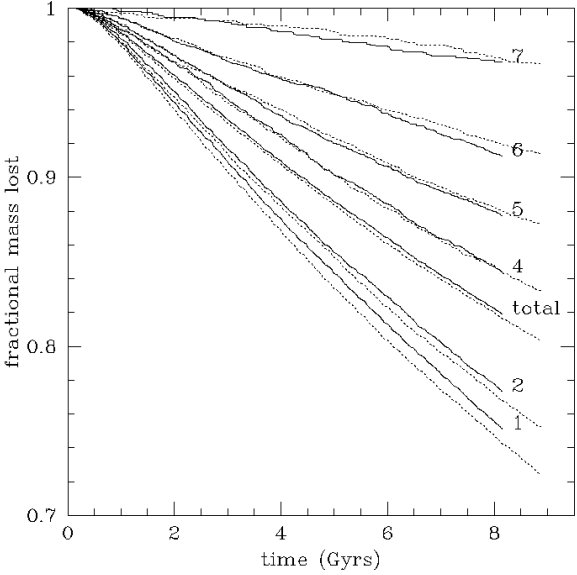}
\caption{Bound mass fraction as a function of time for the labelled mass
groups in Model (0a,p3) run with (dotted lines) and 
without (solid lines) diffusion.
\label{mrelaxfig}}
\end{center}
\end{figure}
{\begin{figure}
\begin{center}
\epsfxsize=8cm \epsfbox{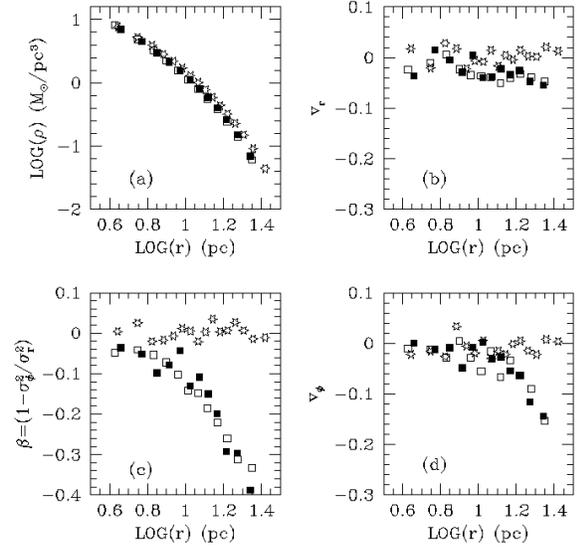}
\caption{Comparison of initial (stars) and final properties
of Model (0a,p3) run with (open squares) and without (closed squares)
diffusion.
\label{vrelaxfig}}
\end{center}
\end{figure}
Figure \ref{mrelaxfig}} compares the mass loss rate for each group in
Model (0a,p3) evolving over 8 Gyrs both with (dotted lines) and
without (solid lines) diffusion included.  The half-mass relaxation
time for this cluster is $T_{relax}\sim 2.68$ Gyrs (see Table
\ref{modstab}).  The evaporation time for a cluster is $\sim$ 100
$T_{\rm relax}$ \cite{me97}, so we expect only a few percent change
in bound mass fraction due to relaxation over the course of the
simulation, as is seen in the figure.  Relaxation contributes to mass
loss by increasing the number of low mass stars lost and decreasing
the number of high mass stars lost, with a net increase in mass loss.
Hence, the evolution of the mass function will be modified from the
purely tidal response described in the previous section.  {Figure
\ref{vrelaxfig}} shows that the density and velocity properties of the
model evolved with (solid squares) and without relaxation (open
squares) are similar (the initial model is shown in stars).

These figures demonstrate that our neglect of relaxation in the models
considered in this paper will not substantially affect their evolution
since a significant fraction of mass is lost due to tidal influences
over a few relaxation times.  In contrast, Pal 5 is an example of a
cluster whose evolution is thought to be dominated by relaxation
\cite{r98} and a natural extension to our work would be to consider
models where the tidal and evaporation mass loss rates are comparable
\cite{vh97}.

%%\clearpage
\section{INTERPRETING OBSERVATIONS OF  
TIDALLY DISRUPTING, MASS-SEGREGATED SYSTEMS}

\subsection{`Observing' the simulations}

Observed properties of globular clusters are often interpreted as
resulting from the tidal influence of the Milky Way.  In this section,
we `observe' our models to confirm the validity of some common
interpretations and explore other signatures of tidal disruption that
future studies might be sensitive to.  We focus our analysis on Model
(0a,p3) as an example of a cluster with a steady mass loss rate that
could survive for the lifetime of the Galaxy.  We examine the
observable properties of this model at the end of the simulation
(after 8 Gyrs of evolution), and compare and contrast it with the same
model at earlier times and with the other models in our sample.

Since the Sun's position along the Solar Circle in the simulations is
arbitrary, rather than assuming a single viewpoint we examine each set
of particle data along three perpendicular axes defined by the
cluster's instantaneous position and velocity: the $x$ axis lies along
the vector joining the cluster to the Galactic centre; the $z$ axis is
perpendicular to the cluster's position and velocity vector (i.e.
looking down on the orbital plane); and the $y$ axis is perpendicular
to these two, along the orbit of the satellite (i.e. it coincides with
the velocity vector when the Galactocentric radial velocity is zero).
We also make the simplification that the lines of sight across the
face of the cluster are perfectly parallel (i.e. the observer is at
infinite distance from the cluster) and state distances in our
projected coordinates in $pc$ rather than $arcmin$. This will not
significantly affect our analysis of density profiles in \S 4.2, but
can alter the measurement of cluster rotation, and we discuss this
simplification further in \S 4.3.

\subsection{Star count profiles}

\subsubsection{Qualitative interpretation}

\begin{figure}
\begin{center}
\epsfxsize=8cm \epsfbox{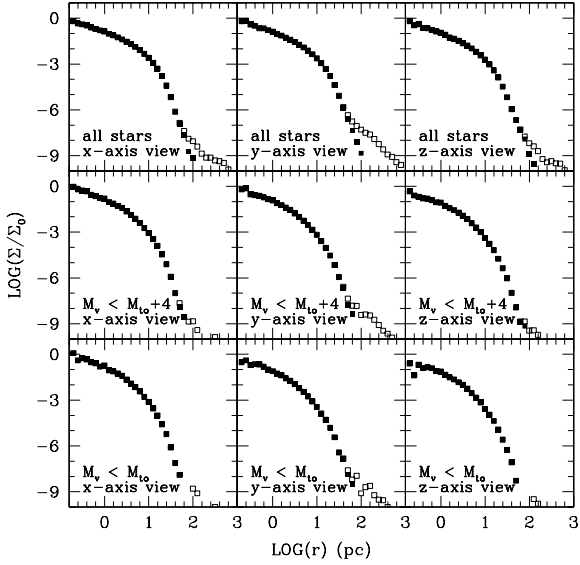}
\caption{Number count
surface density profiles along labelled views (see \S 4.1)
and down to various limiting magnitudes ($M_{\rm to}$ is the 
magnitude of the turnoff)
for the final bound population (filled squares)
and all stars (open squares) for Model (0a,p3).
\label{prof20fig}}
\end{center}
\end{figure}
{Figure \ref{prof20fig}} shows the number surface density for all
stars (top panels), stars brighter than 4 magnitudes below the turnoff
(middle panels) and stars down to the turnoff (bottom panels) at the
end of the simulation of Model (0a,p3) with the three labelled
viewpoints along the axes defined in \S 4.1.  The closed symbols are
for stars still bound to the satellite and the open symbols include
unbound stars in the calculation (where the `bound' and `unbound'
populations are separated following the method described in \S 2.5.1).

\begin{figure}
\begin{center}
\epsfxsize=8cm \epsfbox{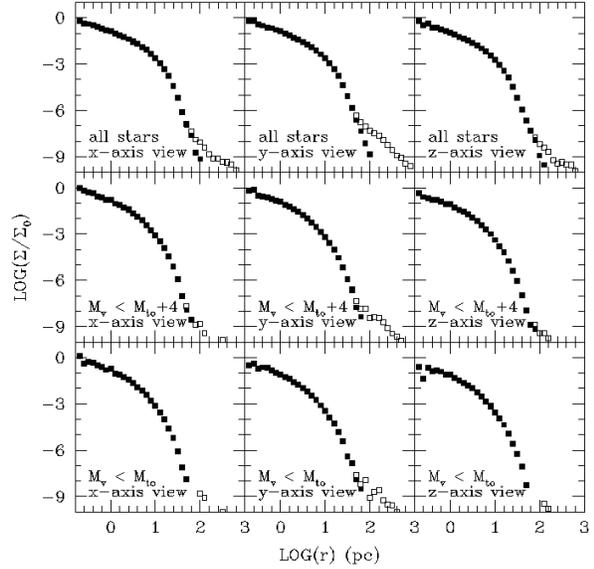}
\caption{Number count
surface density profiles along the $x$-axis for the 
bound population (filled squares)
and all stars (open squares) at random points along
the orbit for labelled
models. The dashed lines show the surface density for
extra-tidal material from equation (\ref{sigma}) and the dotted lines
show the adopted $r_{\rm break}$ in each case, for use in the mass loss
estimates.
\label{proffig}}
\end{center}
\end{figure}
In this Figure, there is a clear break in the slope of the profile
defined by the open symbols, approximately corresponding to the point
where the closed and open symbols become distinct and the fraction of
unbound stars in an annulus becomes significant.  Such `extra-tidal'
stars have been detected in several studies of dwarf spheroidal
galaxies \cite{ih95,ksh96} and globular clusters
\cite{g95,g98}.  Our investigation confirms the common
interpretation that these can be identified as stars escaping from the
satellites.  {Figure \ref{proffig}} demonstrates that this is a
general result by repeating the $x$-axis view for all stars in Model
(0a,3p) at earlier times (left hand panels) and at random points for
several of the other models (as labelled).

\subsubsection{Quantitative interpretation}

Tremaine (1993) pointed out that the change in the orbital frequency of a
star torn from the satellite at radius $r_{\rm break}$ (the point at
which the slope of the surface density profile changes) should
approximately be given by $\Delta \Omega \sim r_{\rm break}
d\Omega/dR$, where $\Omega(R)$ is the frequency of a circular orbit in
the parent galaxy at radius $R$.  Equivalently, the orbital energies
in tidal debris from a satellite orbiting in a potential $\Phi$ will
be spread over a characteristic range $\epsilon=r_{\rm tide}
d\Phi/dR$.  Johnston (1998) tested this simple physical argument by
looking at the spread in energies in streamers seen in numerical
simulations of tidal disruption and found that $\epsilon < |\Delta E|
< 2\epsilon$.  Hence, debris will spread over an angular distance
comparable to the size of the cluster ($r_{\rm break}/R$) in a time
$(r_{\rm break}/R)/2 \Delta \Omega \sim T_{orb}/\pi$, where $T_{\rm
orb}$ is the azimuthal time period of the orbit.  This suggests that
we can estimate the average surface density of stars in an annulus 
between $r_{\rm break}$ and $r$ from the centre of the cluster to be
\begin{eqnarray}
\label{sigbar}
	\langle\Sigma_{\rm xt}(r)\rangle= 
{1 \over g(\theta)}  \left[ {dm \over dt} 
	{(r -r_{\rm break}) \over r_{\rm break}} {T_{\rm orb}
\over \pi}\right] \\
	 \left/ \left[\pi (r^2-r_{\rm break}^2)\right]\right., \nonumber
\end{eqnarray}
where $dm/dt$ is the mass loss rate from the cluster.  The function
$g$ depends on the angle $\theta$ of our line of sight with the plane
perpendicular to the direction of motion of the satellite.
Eventually, the debris spreads out along the satellite's orbit so we
can take $g(\theta)=\cos \theta$ as a crude approximation.  In fact,
mass tends to leave the satellite at the inner and outer Lagrange
points along the vector joining the satellite to the centre of the
galaxy, so the geometry of the streamers close to the satellite is
rather more complicated than this.

We can differentiate equation (\ref{sigbar}) to find an approximate
expression for the absolute surface density
\begin{equation}
\label{sigma}
	\Sigma_{\rm xt}(r) = {1 \over g(\theta)}
{dm \over dt} {T_{\rm orb} \over \pi} {1 \over 2 \pi
r_{\rm break} r}.
\end{equation}
This estimate is overlaid in dashed lines on the profiles shown in
Figure \ref{proffig} using $dm/dt$ averaged over each simulation,
$T_{\rm orb}$ from Table \ref{orbstab} and $r_{\rm break}$ indicated by
the vertical dotted lines in each panel.  The open squares and dashed
line agree well in all cases except for Model (3,p1). In this model,
the simulation was run for only a few orbital periods so the debris
only had time to disperse a few $r_{\rm break}$ from the satellite and
the streamers were not well populated.

Equation (\ref{sigma}) and Figure \ref{proffig} demonstrate that we
expect the extra-tidal population around a cluster to have
$\Sigma_{\rm xt}(r) \propto r^{-1}$.  In observations of real
globulars, the surface density of extra-tidal stars has been found to
fall as $\Sigma_{\rm xt}(r) \sim r^{-\gamma}$ with $-5<\gamma<-0.7$
(G95; 
Zaggia, Piotto \& Capaccioli 1997).  This suggests that the measured $\gamma$
might be used as a test of whether the globular is sufficiently
obscured, either by tidal debris further along its own tidal tail or
by the Galactic field, that the above interpretation is invalid.

Note that these estimates are independent of the method of mass loss
from the cluster since the dynamics once the star is lost is
determined only by the external field, and hence should apply equally
to mass lost by tidal stripping, shocking, or evaporation due to
relaxation.

\subsubsection{Estimating mass
loss rates using extra-tidal stars}

Equations (\ref{sigbar}) and (\ref{sigma}) suggest two ways of
estimating the fractional mass loss rate $(df/dt)$ from a cluster
using observations of extra-tidal stars.

\begin{figure}
\begin{center}
\epsfxsize=8cm \epsfbox{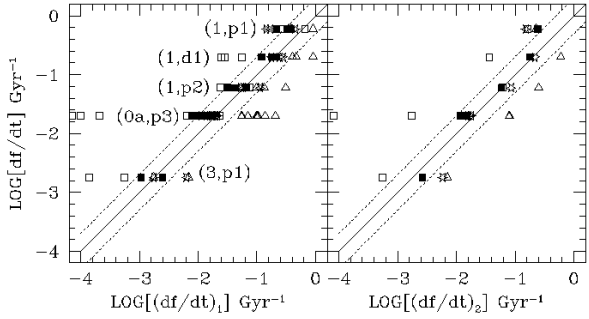}
\caption{Known mass loss rate $df/dt$ (from Figures \ref{masspfig} and
\ref{massdfig}) for the labelled models
plotted against the estimates $(df/dt)_1$
(left hand panel) and $(df/dt)_2$ (right hand panel).
The solid squares, open squares and stars show
the results when viewed from along the $x$, $y$ and $z$-axis
respectively. 
The open triangles repeat the $y$-axis view but with
$g(\theta)\equiv 1$.
The solid line shows where the
estimated and known mass loss rates agree, and the dotted lines are
for factor of 2 discrepancies.
\label{mlossfig}}
\end{center}
\end{figure}
If the extra-tidal population is well defined out to a radius $r_{\rm
xt}$ with $\Sigma_{\rm xt} \sim r^{-1}$ we can count the number of
stars $n_{\rm break}$ within $r_{\rm break}$ (the point where there is
a break in the slope of the surface density profile) and the number of
extra-tidal stars $n_{\rm xt}$ between $r_{\rm break}$ and $r_{\rm
xt}$ and use them to find
\begin{equation}
	\left({df \over dt}\right)_1
= g(\theta) {r_{\rm break}\over r_{\rm xt}-r_{\rm break}}
		{n_{\rm xt} \over n_{\rm break}} {\pi \over T_{\rm orb}}.
\label{dfdt1}
\end{equation}
Since the timescale for stars to diffuse beyond a few $r_{\rm break}$
is an orbital timescale, so long as $r_{\rm xt} > 2 r_{\rm break}$
this estimate should be sensitive to the average mass loss rate rather
than the instantaneous one.  The left hand panel of {Figure
\ref{mlossfig}} plots the known fractional mass loss rate, $df/dt$,
for each of the models with fractional mass loss rates less than
1Gyr$^{-1}$ (see Figures \ref{masspfig} and \ref{massdfig}) against
$(df/dt)_1$ calculated from equation (\ref{dfdt1}).  To find
$(df/dt)_1$, we evaluated the ratio $n_{\rm xt}/n_{\rm break}$ for
$r_{\rm xt}=2 r_{\rm break}$, $r_{\rm xt}= 5 r_{\rm break}$ and
$r_{\rm xt}=10 r_{\rm break}$.  This process was repeated looking
along the $x$-axis (solid squares), $y$-axis (open squares) and
$z$-axis (stars).  The open triangles show the estimate made along
$y$-axis if the viewing angle is not taken into account (i.e.
$g(\theta)\equiv 1$).  The solid line in the figure shows where the
estimated and known mass loss rates agree, and the dotted lines are
for factor of 2 discrepancies.  The figure suggests that we can use
this technique to estimate the mass loss rate from a Galactic
satellite provided our viewpoint lies close to the plane perpendicular
to the satellite's velocity vector (equivalent to the $xz$-plane in
our projection).  Otherwise, we are likely to poorly estimate the rate
of mass loss as the debris along the satellite's orbit confuses the
calculation.

If the extra-tidal population is either not well defined or
non-existent we can estimate $(df/dt)_2$ as an upper limit for the
mass loss rate from $\Sigma_{\rm xt}(r_{\rm break})$, where $r_{\rm
break}$ is taken to be the point either where the slope of the profile
changes or the last point where the surface density is separable from
the background.  Then, from equation (\ref{sigma}),
\begin{equation}
	\left({df \over dt}\right)_{2}=g(\theta)
{\Sigma_{\rm xt}(r_{\rm break})
\over n_{\rm break}} {\pi \over T_{\rm orb}} 2 \pi r^2_{\rm break},
\label{dfdt2}
\end{equation}
which is shown in the right hand panel of Figure \ref{mlossfig}.  As
with the first case, the view perpendicular to the velocity vector
provides the best estimate.
 
\subsection{Line of sight velocity distributions}

\begin{figure}
\begin{center}
\epsfxsize=8cm \epsfbox{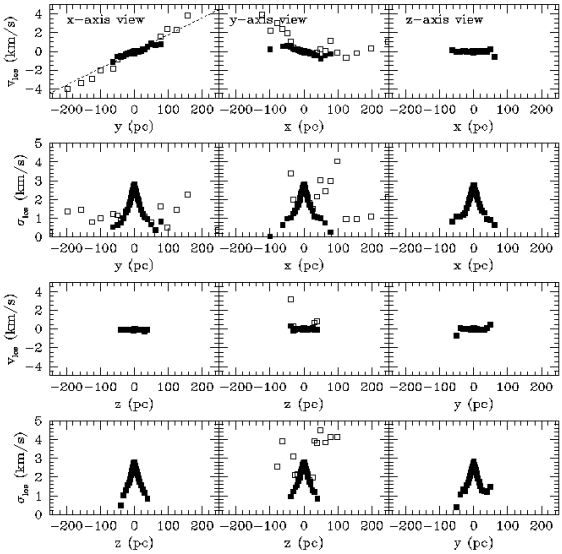}
\caption{Line-of-sight velocity $v_{\rm los}$ and dispersion 
$\sigma_{\rm los}$ for
bound (filled squares) and all stars (open squares) in Model (0a,p3).
The line of sight for each column of panels 
is labelled in the top row (see \S 4.1 for axis definitions).
\label{strip20fig}}
\end{center}
\end{figure}
\begin{figure}
\begin{center}
\epsfxsize=8cm \epsfbox{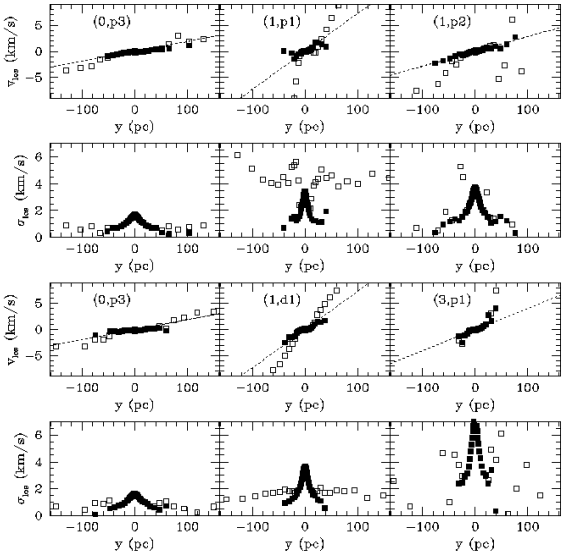}
\caption{Repetition of panels $x$-axis view of Figure \ref{strip20fig}
for the labelled models at random points along the orbit.
\label{stripfig}} 
\end{center}
\end{figure}
{Figure \ref{strip20fig}} summarises the line-of-sight velocity
analysis at the end of the simulation for Model (0a,p3) from the
stated viewpoints and with the observer assumed to be at infinite
distance from the cluster (i.e with the individual lines of sight
across the cluster assumed to be parallel to the one to the cluster
centre -- the effect of this simplification is discussed in \S
4.3.1).  For each viewpoint the average $v_{\rm los}$ and dispersion
$\sigma_{\rm los}$ in the velocities of stars are calculated in bins
lying in strips along the two perpendicular axes across the cluster.
{Figure \ref{stripfig}} demonstrates the general nature of these
plots, repeating the $x$-axis view of Figure \ref{strip20fig} at
different orbital phases of this model and for several different
models.  The solid symbols show the analysis for the bound stars, and
the open symbols shows the results for all stars.

\subsubsection{Average velocities and cluster rotation}

The average velocities are zero except for the viewpoints which are
sensitive to the cluster's orbital motion.  In the panels
corresponding to the $x$-axis view of Figures \ref{strip20fig} and
\ref{stripfig}, the dotted lines indicate the expected velocity
gradient if the cluster were rigidly co-rotating with its orbit.  The
bound stars follow this line fairly closely, with some indication of a
smaller gradient in velocities towards the centres of the clusters.
The unbound stars appear to be rotating faster than this as they move
to orbits with higher/lower angular velocities to form the
leading/trailing streamers.

In our analysis, two effects are clearly contributing to the measured
rotation of the cluster -- the intrinsic rotation of the cluster (in
our simulations, roughly corresponding to co-rotation with the tidal
field), and the velocity gradient in stripped material.  Observers
also have to contend with `perspective rotation' \cite{ftw61,mmm97,d97}
from the projection of the tangential
velocity onto the non-parallel lines of sight across the cluster.  If
we were indeed viewing a rigidly co-rotating cluster from the centre of
the Galaxy, the intrinsic and perspective rotation would cancel each
other out and we would be sensitive only to tidal stripping, but if we
were closer to/farther from the cluster than the centre of the Galaxy,
we would be dominated by the perspective/intrinsic rotation.

Unfortunately, we cannot assume that tidal torquing of a real
satellite necessarily results in co-rotation.  The rotation seen in our
simulations is likely to be an artifact of the near-circular orbits we
have chosen.  In models, run along eccentric orbits, of the disruption
of the Sagittarius dwarf galaxy, Johnston, Spergel \& Hernquist (1995)
found that the line of sight velocity gradient is dominated by
perspective rotation, despite the fact that our viewpoint is farther
from the satellite than the centre of the Galaxy, presumably because
the satellite is not co-rotating. Vel\'azquez \& White (1995) 
made an estimate of
225km s$^{-1}$ for the tangential velocity of the Sagittarius dwarf galaxy by
considering only the effect of perspective rotation and this roughly
agrees with the observed proper motion of 250km s$^{-1}$ \cite{i97}.

Rotation has been detected in several observational surveys of
Galactic globular clusters.  Merritt, Meylan \& Mayor (1997) analysed
the phase-space distribution of 469 stars in $\omega$Cen and found,
after taking perspective rotation into account, it to be consistent
with solid body rotation out to 11pc from the centre of the cluster,
and decreasing beyond.  Drukier et al. (1997), in an analysis of the
velocities of 230 stars in the outskirts of M15, found some evidence
for rotation.  In both these studies, rotation was found to be less
important towards the edge of the cluster, suggesting that in these
cases it is intrinsic and not due to tidal torquing or stripping of
the bound system.

\subsubsection{Velocity dispersions and tidal heating}

In most panels of Figures \ref{strip20fig} and \ref{stripfig} there is
good agreement between $\sigma_{\rm los}$ calculated with and without
the unbound stars.  However, the observed results can be confusing
when looking along the orbit ($y$-axis view panels of Figure
\ref{strip20fig}), as there is significant unbound material along the
line of sight from the trailing and leading streamers.  This problem
may not be so severe in reality if the sample can be selected to
exclude stars beyond a few tidal radii from the cluster (in the
figure, all stars at the projected separation were included).

An observer would also see an apparently enhanced velocity dispersion
when looking at the outskirts of the cluster from the centre of the
Galaxy.  In particular, note that the dispersion in the unbound
material roughly corresponds to the dispersion of the bound material
within the point where stripping occurs.  This suggests that
satellites that are being more violently stripped will have larger
dispersions in their debris trails, though not in excess of the
maximum dispersion of the bound material.  The Sagittarius dwarf
spheroidal galaxy is an example of a system where it can be plausibly
argued that this behaviour is observed: its highly distorted surface
density contours suggest that it is likely to be surrounded by a cloud
of unbound stars and its velocity dispersion is roughly constant along
the entire length of its major axis \cite{i97}.

In their analysis of the velocities in the outskirts of M15, Drukier
et al. (1997) found that the dispersion of the stars decreased to a
minimum at 7 arcmin, increasing again beyond this radius.  They
suggest that this deviation from the behaviour expected for an isolated
cluster could be due to tidal heating, and our simulations confirm
this interpretation.  They comment that the radial position of this
minimum is much smaller than the tidal radius found by G95 by
fitting King Models to star count profiles.  This is also consistent
with our analyses -- the star counts and velocity analyses become
contaminated by unbound stars well within the outermost radius of the
bound system.  Hence the minimum in the velocity dispersion profile is
a good indicator of where this contamination becomes important, but
does not necessarily correspond to the edge of the system.

\subsection{Mass functions}

\subsubsection{Measuring present day mass functions}

\begin{figure}
\begin{center}
\epsfxsize=8cm \epsfbox{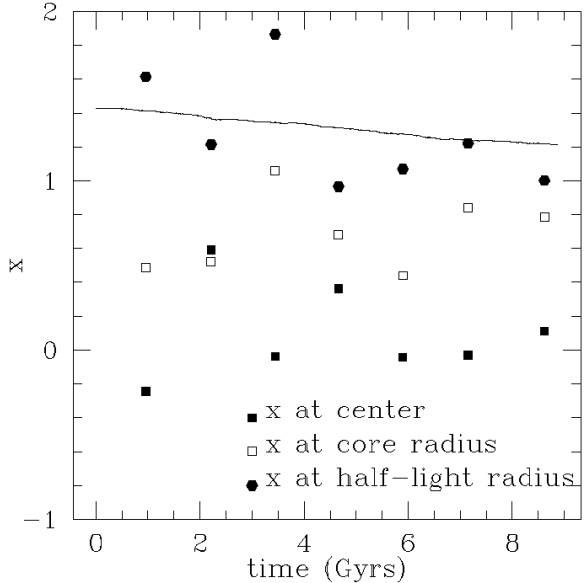}
\caption{Mass function index $x$ as a function of time for all stars
bound to Model (0a,p3) (solid line), and as measured
at various points in the cluster.
\label{pdmffig}}
\end{center}
\end{figure}
\begin{figure}
\begin{center}
\epsfxsize=8cm \epsfbox{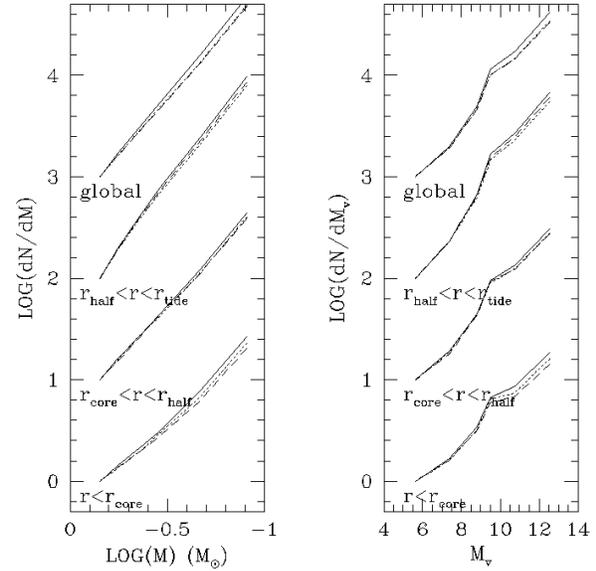}
\caption{Initial (solid lines) and final mass and luminosity functions for
Model (0a:p3) simulated with (dashed lines) and without (dotted lines) 
diffusion effects.
\label{pdmfm0o7fig}}
\end{center}
\end{figure}
In {Figures \ref{pdmffig} and \ref{pdmfm0o7fig}} we mimic the
observational analyses recently applied to the globular clusters NGC
1261 \cite{z98} and M55 \cite{zpc97}.  The solid line in Figure
\ref{pdmffig} shows the evolution of the mass function index $x$ for
Model (0a,p3).  The points indicate what $x$ would be measured to be
at different points along the orbit and at different positions in the
cluster if viewed from the centre of the Galaxy -- within the core
radius (filled squares), between the core and half-light radii (open
squares) and between the half-light and tidal radii (stars).  Each of
these points is calculated using several thousand stars.  In Figure
\ref{pdmfm0o7fig} we plot the mass and luminosity functions at the
beginning (solid lines), and the end (dotted lines) of the simulation.
The dashed lines show the final analysis for the same Model but for
the simulation that included relaxation effects.

\subsubsection{Placing limits on the initial mass function of a cluster}

\begin{figure}
\begin{center}
\epsfxsize=8cm \epsfbox{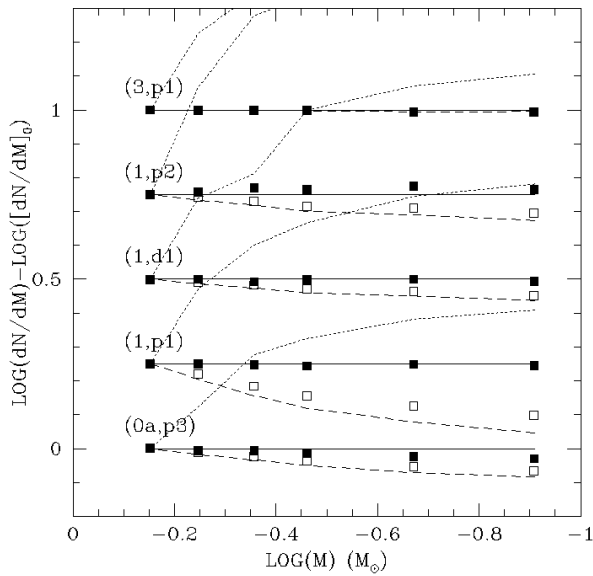}
\caption{Global PDMF (dashed lines) and local PDMF (dotted lines) for outer
100 stars, for the
labelled models. The mass functions have been 
normalized to the IMF of each model
(represented by solid horizontal lines)
and multiplied by arbitrary constants. The solid squares
show the IMF reconstructed
from the estimated mass loss rate, global and
local mass functions.
The open squares show the result of the same 
reconstruction, but using the outermost
1000 stars to find the local PDMF.
\label{imffig}}
\end{center}
\end{figure}
Suppose we have estimated the fractional mass-loss rate from a cluster
using its observed population of extra-tidal stars.  If we also know
the PDMF, both globally and locally in the outskirts of the cluster,
we can place some limits on its IMF.  We test this idea with our
simulations by using the methods outlined in \S 4.1 to find
$(df/dt)_1$.  {Figure \ref{imffig}} shows the final mass function
$(dN/dM)_{\rm final}$ at the end of the five labelled simulations
(dashed line) and the mass function for the 100 stars at the projected
edge of each system $(dN/dM)_{\rm edge}$ (dotted line), where each
mass function has been scaled by the known IMF ($(dN/dM)_0$,
represented by the horizontal solid line).  We then simply estimate
the IMF (solid squares) to be,
\begin{equation}
	\left({dN \over dM} \right)_{0, \rm est}=
\left({dN \over dM} \right)_{\rm final}+
T_{\rm sim} \left({df \over dt}\right)_{1}
\left({dN \over dM} \right)_{\rm edge}
\label{imf}
\end{equation}
where $T_{\rm sim}$ is the duration of the simulation.  (The open
squares show the result if the PDMF of the outermost 1000 stars is
used.)

Several assumptions make this estimate uncertain: (i) $(df/dt)_{1}$ is
taken from star counts, implicitly assuming that the mass function of
the extra-tidal material is the same as the global mass function; (ii)
we assume the mass-function of the stripped material to be constant
with time; (iii) we neglect relaxation effects; and (iv) we assume
that the satellite fills its tidal radius, and hence loses mass at a
roughly constant rate due to either evaporation or tidal effects.  The
first approximation can be addressed if the luminosity function of the
cluster is known as a function of radius, as is the case in our
simulations -- Figure \ref{imffig} demonstrates that we can make
reasonable predictions for the simulations even without this level of
refinement; the second two assumptions will both lead to an
underestimate of the evolution of IMF; and the last approximation will
place some limits on how far back we might be able to `integrate'
the differential mass loss.

Despite these uncertainties, this method provides a new approach --
more directly based on observations rather than using complex
dynamical models -- to exploring the question of whether the IMF in
globulars was universal, or environment dependent.

%%\clearpage
\section{DISCUSSION: APPLICATIONS TO OBSERVED SYSTEMS}

\subsection{Destruction rate of galactic satellites}

\subsubsection{Dwarf spheroidal galaxies}

IH95 studied the morphology of eight of the dwarf spheroidal
satellites of the Milky Way (all the ones currently known with the
exception of Sagittarius) determined from star counts made using the
APM facility at Cambridge.  Extra-tidal stars are clearly seen in the
number count profiles for six of these satellites, but in no case do
they clearly follow $\Sigma_{\rm xt} \sim r^{-1}$.  Hence, we use
equation (\ref{dfdt2}), to find an upper limit for the mass loss rate.
Using the data from Table 3 of IH95 (kindly made available to
us by M. Irwin), we subtract off the mean background stellar density
from each of their bins and take $r_{\rm break}$ to be the smaller of
the point at which the slope of the star count profile changes (by
visual inspection of their figure 2) and the last measured surface
density above the background.  The orbital time period $T_{\rm orb}$
is calculated for a circular orbit in a logarithmic potential, with
circular velocity $v_c=200$km s$^{-1}$
\begin{equation}
\label{torb}
	T_{\rm circ}={2 \pi R_{\rm GC} \over v_c}=
		  1.5 \left({R_{\rm GC} \over 50 {\rm kpc}}\right) {\rm Gyrs},
\end{equation}
with $R_{\rm GC}$ taken to be the current distance of each satellite
from the Galactic centre.  In fact, if the time period of orbits is
assumed to be independent of angular momentum (shown to be
approximately true in Johnston 1998), then a satellite at $R_{\rm GC}$
with velocity $v$ in the range $0< v < 3 v_{\rm c}/2$ will have an
orbital time period $T_{\rm orb}$ in the range $T_{\rm circ}(R_{\rm
GC})/2 < T_{\rm orb} < 2 T_{\rm circ}(R_{\rm GC})$, so equation
(\ref{torb}) is expected to be good to within a factor of 2.  Several
of the dwarf spheroidal satellites do have proper motion measurements
-- with large errors. These could in principle be used to determine a
specific time period, but since the mass loss estimate itself is
expected to be uncertain to within a factor of 2, the approximation in
equation (\ref{torb}) is deemed to be sufficiently accurate.  For
those with proper motions we calculate the angle $\theta$ between our
line of sight and the plane perpendicular to the velocity vector of
the satellite in the Galactic rest frame.  Otherwise, we assume the
orbit is circular and calculate $\theta$ as the angle between our line
of sight and the Galactocentric radius vector of the satellite.

\begin{table*}
\begin{minipage}{16cm}
\caption{ Mass loss rates for dwarf spheroidals.}
\label{dsphtab}
\begin{tabular}{|ccccc|}
\hline
Name    & $R_{\rm GC}$  & $T_{\rm orb}$ & $\theta$      & $(df/dt)_2$ \\
        & (kpc)         & (Gyr)         & (degrees)     & (Gyr$^{-1}$) \\
\hline
\hline
Carina  & 8.66E+01      & 2.60E+00      & 2.35E+01      &  $<$ 3.31E-01 \\

Draco   & 7.20E+01      & 2.16E+00  & 4.50E+01  &  $<$ 2.18E-01 \\

Fornax  & 1.22E+02 &  3.66E+00  & 2.94E+01  &  $<$ 6.17E-02 \\

LeoI  & 2.02E+02  & 6.05E+00  & 5.61E+01  &  $<$ 5.92E-02 \\

LeoII  & 2.10E+02  & 6.29E+00  & 3.66E+01  &  $<$ 1.44E-01 \\

Sculptor  & 7.22E+01  & 2.16E+00  & 1.64E+01  &  $<$ 2.84E-01 \\

Sextans  & 8.60E+01  & 2.58E+00  & 4.36E+01  &  $<$ 2.61E-01 \\

Ursa Minor  & 6.59E+01  & 1.98E+00  & 1.30E+01  &  $<$ 3.22E-01 \\
\hline
\end{tabular}

\medskip
Columns: (1) name; (2) Galactocentric distance; (3) time period of circular
orbit at that distance; (4) angle between line of sight and
plane perpendicular to satellite's velocity,
calculated from proper motion measurements for
Sculptor \cite{s95} and Ursa Minor \cite{s98}; 
(5) mass loss  rate estimate
from equation (\ref{dfdt2}).
\end{minipage}
\end{table*}
All these quantities are shown in {Table \ref{dsphtab}}.  The
estimates $(df/dt)_2$ range from a few percent (Fornax and Leo I) up
to more than 30 percent in the next Gyr (Ursa Minor and Carina).  They do
not correlate well with previous calculations that attempted to
determine the robustness of each object either from the ratio of the
expected to observed tidal radius (where the former was calculated
given a dynamical estimate of the satellite's mass), or of the
external tidal field to internal field (see IH95).  Nevertheless,
these mass loss rates clearly indicate that the satellite system of
the Milky Way could easily be diminished by several members in the
next 10 Gyrs, which in turn suggests that there may have been several
more satellites of the Milky Way in the past.

If these dSph do indeed have such large mass loss rates, why have we
not detected tidal streamers?  In the cases of Ursa Minor and Sculptor
this question was addressed by Johnston (1998), using a semi-analytic
technique.  She found that if each was losing mass at the rate of 10 percent
per Gyr the local number count densities along the streamers (i.e. not
averaged over annuli centred in the cluster) would not exceed 1 percent of
the background star counts predicted by the Bahcall-Soneira 
\cite{bs80} model of the Milky Way.  Hence, even increasing these
mass loss rates to the tabulated values would not make the streamers
striking features in the sky.  However, the estimates bolster the notion
that these streamers might be discovered and traced over large angular
extents using integrated star counts (such as the method of Great
Circle Cell Counts proposed by Johnston, Hernquist \& Bolte 1996) 
or with color and
velocity information to distinguish them from the background.

\subsubsection{Globular clusters}

\begin{table*}
\begin{minipage}{16cm}
\caption{ Mass loss rates for globular clusters.}
\label{globstab}
\begin{tabular}{|cccccccc|}
\hline
NGC      &  $R_{\rm GC}$ & $T_{\rm orb}$ & $\theta$ & $\gamma$ & $(df/dt)_1$ & 
$(df/dt)_2$ & G\&O \\ 
         & (kpc)     & (Gyr)    & (degrees) & & (Gyr$^{-1}$) & (Gyr$^{-1}$) 
& (Gyr$^{-1}$)\\
\hline
\hline
   288 & 1.14E+01 & 3.43E-01 & 4.19E+00 & -5.21E-01 & 5.26E-02  & 1.26E-02 & 
1.10E-01\\

   362 & 9.04E+00 & 2.71E-01 & 1.40E+01 & -1.58E-01  & 6.22E-01  & 5.77E-01 & 
3.54E-02 \\

  1904 & 1.81E+01 & 5.44E-01 & 8.66E+01 & --  &--  & $<$ 8.51E-03 & 3.54E-02 
\\

  2808 & 1.07E+01 & 3.20E-01 & 2.33E+01 & --  &-- & $<$ 4.37E-02 & 1.61E-02 \\

  3201 & 8.85E+00 & 2.65E-01 & 4.95E+01 & --  &-- & $<$ 4.50E-01 & 3.45E-02 \\

  4590 & 9.94E+00 & 2.98E-01 & 5.17E-01 & -- & --  & $<$ 1.28E+00 & 8.22E-03 
\\

  5824 & 2.60E+01 & 7.81E-01 & 7.94E+01 &-1.33E+00  & 6.46E-02  &1.18E-01 & 
3.06E-03\\

  6864 & 1.17E+01 & 3.52E-01 & 8.60E+01 & -- & --  & $<$ 4.91E-02 &  1.89E-02 
\\

  6934 & 1.17E+01 & 3.52E-01 & 4.33E+01 & --  &--  & $<$ 4.18E-01 & 2.89E-02 
\\

  6981 & 1.22E+01 & 3.65E-01 & 5.84E+01 & --  &-- & $<$ 2.96E-01 & 1.76E-02 \\

  7078 & 1.01E+01 & 3.02E-01 & 5.06E+01 & --  &--  & $<$ 7.00E-02 & 2.17E-02 
\\

  7089 & 1.01E+01 & 3.02E-01 & 1.91E+01 &-1.62E+00  &1.29E-01 & 3.13E-01 & 
5.76-03 \\
\hline 
\end{tabular}

\medskip
Columns: (1) name; (2) Galactocentric distance; (3) time period of circular
orbit at that distance; (4) angle between line of sight and
plane perpendicular to satellite's velocity,
calculated from proper motions in Dauphole et al. 1996;
(5) slope of extra-tidal star
surface density profile; (6) mass loss rate estimate
from equation (\ref{dfdt1});
(7) mass loss  rate estimate
from equation (\ref{dfdt2})
(8) mass loss rate estimate from Gnedin \& Ostriker (1997).
\end{minipage}
\end{table*}
{Table \ref{globstab}} presents the results of mass loss estimation for
the 12 globular clusters analysed by G95 (using their tables
3-14, kindly made available to us by C. Grillmair).  Four of these
clusters have well-defined tidal tails with slopes $\gamma \approx 1$
(given in column 5), and in these cases both $(df/dt)_1$ and
$(df/dt)_2$ are calculated.  In the other cases, only the latter
estimate is made as an upper limit on the mass loss rate.  The
estimated limits for the mass loss rates range from a few to over 100
percent in the next Gyr, again implying that the Galaxy's globular
cluster system will evolve substantially in the next Hubble time.
This provides observational support for the many purely theoretical
studies that have reached the same conclusion using semi-analytic
models \cite{aho88,go97,mw97,v97,ch96}.

The last column shows the destruction rates predicted by 
Gnedin \&\ Ostriker (1997).
There is no striking correspondence between our
`observational' results and the purely `theoretical' values.
However, the upper limit we calculate is only in direct contradiction
with the theoretical calculation in the cases of NGC 288 and NGC 1904.
In the former case our viewing angle is favourable for making an
accurate estimate, but in the latter the value of $\theta$ is
sufficiently large that we might expect the result to be confused by
the debris geometry.  In general, our mass loss rates are higher, but
Gnedin \&\ Ostriker (1997)
themselves point out that their destruction
rates should be taken as a lower limit since the rates could be
increased with the inclusion of other effects ignored in their study
(such as a mass-spectrum).

\subsubsection{Future prospects}

The results in Tables \ref{dsphtab} and \ref{globstab} should be treated
with some caution as in most cases the observed densities of
extra-tidal stars do not follow $\Sigma(r) \propto r^{-1}$ predicted
by the simple model which we use to interpret the data.  However, the
original profiles were made either using star counts directly, or with
additional photometry to subtract off some of the background.  Since
we expect the tidal debris to typically have velocities within $\pm 10
{\rm \, km\, s^{-1}}$ of the satellite, 
a study of radial velocities around a cluster
has the potential of refining these measurements considerably.
Further in the future, the astrometric satellites SIM (the Space
Interferometry Mission) and GAIA (the Galactic Astrometric Imaging
Satellite) promise proper motion measurements with a few $\mu arcsec$
accuracy (or tangential velocities to better than 1 km s$^{-1}$ out to tens
of kpc), and a similar advance in identifying debris.

We have also, for the sake of simplicity, restricted our discussion to
annularly averaged surface densities.  Clearly, more information is
contained in two-dimensional surface density maps 
\cite{g95}.  However, these will be more sensitive to the orbital
phase and the mass loss history of the satellite and would require
more detailed analytic modeling to interpret.

\subsection{Determining the IMF from observations of the PDMF}

The method proposed in \S 4.5 uses observations of a cluster's PDMF and
extra-tidal stars to place some limits on the IMF.  This is clearly a
powerful tool for making progress towards understanding whether the
IMF is in any sense `universal'.

Of course it is non-trivial to find the global PDMF, the local mass
function in the exterior, and to detect extra-tidal stars around a
cluster.  However, there are currently two examples in the literature
where this has already been done -- M15 and M55 (see G95;
Piotto, Cool \& King 1997; 
Zaggia et al. 1997).  In the case of M15, Piotto, Cool \& King
(1997) argue that mass-segregation is important only in the centre of
the cluster and that their local sample is a fair representation of
the global PDMF.  Unfortunately, in the absence of mass segregation
our method would find no evolution of the IMF since it does not model
differential mass-loss due to relaxation effects, but only steady
stripping of the most weakly bound stars.  (Since Piotto et al. 1997 
make
the same statement about NGC 6397, which has a very different $x$ from
M15, this might imply that the differences can only be due to
relaxation effects or an intrinsically
different IMF -- it is unclear whether this is a robust
argument, or whether a low level of mass segregation farther out in
the cluster might be sufficient to account for the differences through
tidal shocking.)  In the case of M55, Zaggia, Piotto \& Capaccioli
(1997) find some evidence for extra-tidal stars, but do not give the
density of this material since there are several other effects that it
could be attributed to.  Despite these problems, these studies show
that it is currently feasible to design future observations that can
address these issues.

%%\clearpage
\section{CONCLUSIONS}

We summarise our main conclusions as follows:
\begin{enumerate}
	\item Observations of extra-tidal stars and enhanced velocity
dispersions in the outskirts of clusters can be attributed to tidally
stripped material.  The star counts become dominated by unbound stars
at the point where the slope of the surface density profile changes or
the dispersion reaches a minimum.  This should not be identified with
the tidal radius of the cluster since the edge of the bound population
can still lie significantly beyond this radius.
	\item The mass loss rate from a Galactic satellite can be
directly estimated from the population of extra-tidal stars within a
few of its tidal radii, the orbital time period and our line of sight.
	\item Using current observations we calculated the mass-loss
rate from dwarf spheroidal satellites and globular clusters and found
that both systems will undergo significant evolution in the next
Hubble time.  Future observations should be able to place much
stronger limits on these destruction rates.
	\item Mass-loss estimate together with measurements of the
local and global PDMFs of a cluster can place limits on its IMF.  Our
review of the literature suggests that it is currently observationally
feasible to carry out this program.
\end{enumerate}

\section*{Acknowledgements}

We are grateful to Piet Hut, Douglas Heggie and the other occupants of
E-building at the IAS for helpful discussions, and to the IAS, IoA
Cambridge and Sterrewacht Leiden for hospitality.  We thank Mike Irwin
and Carl Grillmair for providing data used for our analysis.  This
work was supported in part by 
the National Center for Supercomputing
Applications, the Pittsburgh Supercomputer Center, and the EPCC; NASA
Grant NAGW--2422 and the NSF under grants AST 90--18526, ASC
93--18185, the Presidential Faculty Fellows Program; funds from
the IAS, a PPARC theory
grant, an EU Marie Curie Fellowship and the generous support of the
British Council.

%\clearpage


\begin{thebibliography}{}

\bibitem[Aguilar et al. 1988]{aho88} Aguilar, L., Hut, P. \&\
    Ostriker, J.P., 1988, \apj, 335, 720

\bibitem[Bahcall \& Soneira 1980]{bs80}
Bahcall, J. N. \& Soneira, R. M., 1980, \apjs, 44, 73

\bibitem[Bergbusch \& VandenBerg's (1992]){bv92}
Bergbusch, P. A. \& VandenBerg, D. A., 1992, \apjs, 81, 163

\bibitem[Binney \& Tremaine 1987]{bt}
 Binney, J. \& Tremaine, S., 1987, Galactic Dynamics
(Princeton University Press, Princeton)

\bibitem[Capaccioli et al. 1991]{cop91}  
Capaccioli, M., Ortolani, S. \& Piotto, G., 1991, \aap, 244, 298

\bibitem[Capaccioli, Piotto \& Stiavelli 1993]{cps93}  
Capaccioli, M., Piotto, G. \&
Stiavelli, M., 1993, \mnras, 261, 819

\bibitem[Capriotti \& Hawley 1996]{ch96}
Capriotti, E.R. \& Hawley, S.L., 1996, \apj, 464, 765

\bibitem[Chernoff \&\ Weinberg 1990]{cw90}  Chernoff, D.F. \&\
    Weinberg, M.D., 1990, \apj, 351, 121

\bibitem[Da Costa \& Freeman 1976]{dc76}  Da Costa, G.S. \& Freeman, K., 1976,
\apj, 206, 128

\bibitem[Dauphole et al. 1996]{d96}
Dauphole, B., Geffert, M., Colin, J., Ducourant, C.,
Odenkirchen, M. \& Tucholke, H.J., 1996, \aap, 313, 119

\bibitem[Djorgovski, Piotto \& Capaccioli 1993]{dpc93}
Djorgovski, S., Piotto, G. \& Capaccioli, M., 1993, \aj, 105, 2148

\bibitem[Drukier et al. 1997]{d97}
Drukier, G.A., Slavin, S.D., Cohn, H.N., Lugger, P.M., Berrington, R.C.,
Murphy, B.W., \& Seitzer, P.O., 1997, \aj, 115, 708

\bibitem[Elson, Hut \& Inagaki 1987]{el87} Elson, R.A.W., Hut, P. \&\
    Inagaki, S., 1987, \araa, 25, 565

\bibitem[Elson et al. 1998]{e98} Elson, R.A.W., Sigurdsson, S., Davies, M.B.,
     Hurley, J. \& Gilmore, G., 1998, \mnras, in press

\bibitem[Faber \& Lin 1983]{fl83}
Faber, S.M. \& Lin, D.N.C., 1983, \apjlett, 266, L17

\bibitem[Fall \&\ Rees 1977]{fr77}  Fall, M. \&\ Rees, M.J., 1977,
    \mnras, 181, 37P

\bibitem[Feast, Thackeray \& Wesselink, 1961]{ftw61}
Feast, M.W., Thackeray \& Wesselink, A.J., 1961, \mnras, 122, 433 

\bibitem[Fischer et al. 1998]{f98}
Fischer, P., Pryor, C., Murray, S., Mateo, M. \& Richtler, T., \aj, 1998, 115, 
592

%\bibitem[Fukushige \&\ Heggie 1005]{fu95}  Fukushige, T. \&\
%    Heggie, D.C., 1995, \mnras, 276, 206

\bibitem[Gnedin \&\ Ostriker 1997]{go97}  Gnedin, O. \&\
    Ostriker, J.P., 1997, \apj, 488, 579

\bibitem[Gnedin, Hernquist \&\ Ostriker 1998]{gho98}  Gnedin, O.,
    Hernquist, L. \&\ Ostriker, J.P., 1998, \apj, in press

\bibitem[Grillmair 1998]{g98}  
Grillmair, C.J., 1998, in Galactic Halos, ed. D. Zaritsky,
A.S.P. Conf. Ser, 136, 45

\bibitem[G95]{g95}  Grillmair, C.J., Freeman, K.C.,
    Irwin, M. \&\ Quinn, P.J., 1995, \aj, 109, 2553

\bibitem[Gunn \& Griffin 1979]{gg79}  Gunn, J.E. \& Griffin, R.F., 1979,
\aj, 84, 752

\bibitem[Hernquist 1990]{h90} 
Hernquist, L., 1990, \apj, 356, 359

\bibitem[Hernquist \&\ Ostriker 1992]{ho92}  Hernquist, L. \&\
    Ostriker, J.P., 1992, \apj, 386, 375

\bibitem[Hernquist et al. 1995]{her95}  Hernquist, L., Sigurdsson, S.
    \&\ Bryan, G.L., 1995, \apj, 446, 717

\bibitem[Hut \&\ Djorgovski (1992)]{hd92}  Hut, P. \&\ Djorgovski, S.,
    1992, Nature, 359, 806

\bibitem[Ibata et al. 1997]{i97}
Ibata, R. A., Wyse, R.F.G., Gilmore, G., Irwin, M. J. \& Suntzeff, N.B.,
1997, \aj, 113, 634

\bibitem[IH95]{ih95}
Irwin, M. J. \& Hatzidimitriou, D., 1995, \mnras, 277, 1354 (IH95)

\bibitem[Johnston 1998]{j98}
Johnston, K.V., 1998, \apj, 495, 297

\bibitem[Johnston, Spergel  \& Hernquist 1995]{jsh95} 
Johnston, K. V., Spergel, D. N.  \& Hernquist, L., 1995, \apj, 451, 598

\bibitem[Johnston, Hernquist \& Bolte 1996]{jhb96} 
Johnston, K. V., Hernquist, L. \& Bolte, M., 1996, \apj, 465, 278

\bibitem[Johnston et al. 1998]{jhw98} 
Johnston, K. V., Hernquist, L. \& Weinberg, M.D., 1998, \apj,
submitted

\bibitem[King 1962]{k62} 
King, I. R., 1962, \aj, 67, 471

\bibitem[Kuhn et al. 1996]{ksh96}
Kuhn, J. R., Smith, H. A.  \& Hawley, S. L., 1996, \apjlett, 469, L93

\bibitem[Kundic \&\ Ostriker 1995]{ko95}  Kundic, T. \&\
    Ostriker, J.P., 1995, \apj, 438, 702

%\bibitem[Lee \&\ Goodman 1995]{le95}  Lee, X. X. \&\ Goodman, J.
%    1995, \apj, 443, 109

\bibitem[McClure et al. (1986)]{mc86}  McClure, R.D. et al., 1986,
    \apjl, 307, L49

\bibitem[Merritt, Meylan \& Mayor 1997]{mmm97}
Merritt, D., Meylan, G. \& Mayor, M., 1997, \aj, 114, 1074

\bibitem[Meylan \&\ Heggie 1997]{me97}  Meylan, G. \&\ Heggie, D.C.,
    1997, \aapr, 8, 1

\bibitem[Miyamoto  \& Nagai 1975]{mn75} 
Miyamoto, M.  \& Nagai, R., 1975, \pasj, 27, 533

\bibitem[Murali \& Weinberg 1997]{mw97}
Murali, C. \& Weinberg, M. D.  1997, \mnras, 291, 717

\bibitem[Oh, Lin \& Aarseth 1995]{ola95}
Oh, K.S., Lin, D.N.C. \& Aarseth, S.J., 1995, \apj, 442, 142

%\bibitem[Piatek \& Pryor 1995]{pp95} 
%Piatek, S. \& Pryor, C., 1995, \aj, 109, 1071

\bibitem[Piotto et al. 1997]{pck97}
Piotto, G., Cool, A.M. \& King, I.R., 1997, \aj, 113, 1354

%\bibitem[Pryor et al 1986]{pr86}  Pryor, C., Smith, G.H. \&\
%    McClure, R.D., 1985, \aj, 92, 1358

\bibitem[Rosenberg et al. 1998]{r98}
Rosenberg, A., Saviane, I., Piotto, G., Aparicio, A. \& Zaggia, S.R.,
1998, \aj, 115, 658

%\bibitem[Spergel 1995]{s95}
%Spergel, D.N.S., 1995

\bibitem[Sigurdsson \&\ Phinney 1995]{sp95}  Sigurdsson, S. \&\
    Phinney, E.S., 1995, \apjs, 99, 609

\bibitem[Sigurdsson et al. 1997]{ss97} Sigurdsson, S., He, B., Melhem, R.
    Hernquist, L., 1997, Computers in Physics, 11.4, 378

\bibitem[Spitzer 1958]{s58}
Spitzer, L., 1958, \apj, 127, 17 

\bibitem[Schweitzer, Cudworth, Majewski \& Suntzeff (1995)]{s95}
Schweitzer, A.  E., Cudworth, K.  M., Majewski, S. R. \& Suntzeff, N. B.,
1995, \aj, 110, 2747

\bibitem[Schweitzer, Cudworth \& Majewski (1998)]{s98}
Schweitzer, A.  E., Cudworth, K.  M. \&  Majewski, S. R.,
1998, \aj, submitted

\bibitem[Tremaine (1993)]{t93}
Tremaine, S., 1993 in Back to the Galaxy eds. S. S. Holt \& F. Verter,
(AIP Conf. Proc. : New York), p.599

\bibitem[Vel\'azquez \& White (1995)]{vw95}
Vel\'azquez , H. \& White, S.D.M., 1995, \mnras, 275, L23

\bibitem[Vesperini 1997]{v97}
Vesperini, E., 1997, \mnras, 287, 915

\bibitem[Vesperini \& Heggie 1997]{vh97}
Vesperini, E. \& Heggie, D.C., 1997, \mnras, 289, 898

\bibitem[Weinberg 1994]{w94}  Weinberg, M.D., 1994 \apj, 108, 1403

\bibitem[Zaggia et al. 1997]{zpc97}
Zaggia, S.R., Piotto, G. \& Capaccioli, M., 1997, \aap, 327, 1004

\bibitem[Zoccali et al. 1998]{z98}
Zoccali, M., Piotto, G., Zaggia, S.R., \& Capaccioli, M., 1998, 
\aap, 331, 541
\end{thebibliography}
\end{document}